\documentclass{article}

\usepackage{graphicx}
\usepackage{amsmath}
\usepackage{amssymb}
\usepackage[numbers,sort&compress]{natbib}
\usepackage{subfig}
\usepackage{caption}
\usepackage[utf8]{inputenc}
\usepackage{url}

\title{Electromagnetic scattering in Schwarzschild space-time: Finite difference time domain with Green function method}
\author{Shouqing Jia\\School of Computer Science and Engineering, Northeastern University\\Shenyang, 110819, P. R. China\\jiashouqing@163.com}

\begin{document}
\maketitle

\section*{Abstract}

The finite difference time domain (FDTD) algorithm and Green function algorithm are implemented into the numerical simulation of electromagnetic scattering by ordinary objects in Schwarzschild space-time. FDTD method in curved space-time is developed by filling the flat space-time with an equivalent medium. Green function in curved space-time is obtained by solving transport equations. Simulation results validate both the FDTD code and Green function code. Scattering in Schwarzschild space-time is simulated by these methods.

\section*{Keywords}
Electromagnetic scattering, Schwarzschild space-time, Finite difference time domain, Green function

\section{Introduction}
Recently, with the booming of manned space industry and development of deep space communication based on electromagnetic waves among spacecrafts, it becomes more and more importantly to study the properties of the electromagnetic waves in curved space-time.

The propagation of electromagnetic waves in curved space-time has been studied \cite{1,2,3} since the establishment of the theory of general relativity. Numerical simulation of electromagnetic waves in curved space-time also attracted researchers' interest. Daniel and Tajima \cite{4} used the electromagnetic particle-in-cell (EMPIC) algorithm to study the physics of high-frequency electromagnetic waves in a general relativistic plasma with the Schwarzschild metric. Watson and Nishikawa \cite{5} incorporated the Kerr-Schild metric into the EMPIC code for the simulation of charged particles in the region of a spinning black hole. The scattering by black holes has also been studied \cite{bh1,bh2,bh3,bh4}. The scattering of a planar monochromatic electromagnetic wave incident upon a Schwarzschild black hole is analyzed in \cite{bh1}. The absorption cross section of Reissner-Nordström black holes for the electromagnetic field is computed numerically in \cite{bh2}. The Orbiting phenomena in black hole scattering is studied in \cite{bh3}. The wavefront twisting by rotating black holes is studied in \cite{bh4}. However, electromagnetic scattering by ordinary objects (not black holes) in curved space-time were not involved in these studies. The numerical methods to simulate electromagnetic scattering by ordinary objects are developed in this paper. 

Finite difference time domain (FDTD) method \cite{6,7,8} is one of the most popular numerical methods for simulating electromagnetic waves in flat space-time. The FDTD method can be easily extended to curved space-time \cite{4,5}. The iteration formulas are deduced from the 4-D Maxwell equations in \cite{5}. The difference is that, in this paper, we directly use the flat space-time FDTD method, in which the space is filled with an inhomogeneous medium.

In the theory of general relativity, space-time is curved and light propagates along geodesics. It is well recognized that Maxwell's equations in curved space-time can be written as if they were in a flat space-time with an optical medium, which is described by a constitutive equation \cite{9,10,11,12}. From this point of view, theoretical methods for designing devices that offer unprecedented control over electromagnetic fields can be developed. Thus, such methods open up new avenues to design and realize functional electromagnetic devices \cite{13,14,15}. Since the curvature of space-time is equivalent to certain medium, the FDTD method in flat space-time can be directly used. This is introduced in section \ref{sec. fdtd method}. It will be seen that the effective permittivity and permeability in Schwarzschild space-time have very simple expressions.

The FDTD method is simple and straightforward, while to solve electromagnetic scattering problems, the Green function is indispensable. Green function in flat space-time takes very simple form. However, in curved space-time the Green function is rather complex \cite{16,17}. To calculate Green function in curved space-time, differential equations should be solved numerically \cite{18,19}. This is introduced in Section \ref{sec. green function}. The connection boundary and output boundary in FDTD method are introduced in Section \ref{sec. cb ob}. Some numerical results are shown in section \ref{sec. Numerical results}. 
 
\section{FDTD method in Schwarzschild space-time}\label{sec. fdtd method}
In this paper, tensor indices run from 0 to 3. The Minkowski metric is
\begin{equation}
    \eta_{\alpha\beta}=\text{diag}(-1,1,1,1)
\end{equation}
The exterior differential form of Maxwell equations are:
\begin{equation} \label{Maxwell-exterior differential}
    \left\{
        \begin{aligned}
			d*F&=Z_0*J\\
			dF&=*\tilde{J}
        \end{aligned}
    \right.
\end{equation}
where $Z_0$ is the vacuum wave impedance, $F$ is electromagnetic tensor, $J$ and $\tilde{J}$ are electric current density and magnetic current density respectively, $*$ is Hodge star operator, and $d$ is exterior differential operator. The covariant derivative form of Maxwell equations are:
\begin{equation} \label{Maxwell-covariant derivative}
    \left\{
        \begin{aligned}
            &F^{\mu\nu}{}_{;\nu}=Z_0J^\mu\\
            &F_{\nu\sigma;\mu}+F_{\mu\nu;\sigma}+F_{\sigma\mu;\nu}=(*\tilde{J})_{\mu\nu\sigma}
        \end{aligned}
    \right.
\end{equation}
where the semicolon represent covariant derivative. The partial derivative form of Maxwell equations are:
\begin{equation} \label{Maxwell-partial derivative}
    \left\{
        \begin{aligned}
            &\frac{1}{\sqrt{-g}}\frac{\partial}{\partial x^\nu}(\sqrt{-g}F^{\mu\nu})=Z_0J^\mu\\
            &\frac{\partial F_{\nu\sigma}}{\partial x^\mu}+\frac{\partial F_{\mu\nu}}{\partial x^\sigma}+\frac{\partial F_{\sigma\mu}}{\partial x^\nu}=(*\tilde{J})_{\mu\nu\sigma}
        \end{aligned}
    \right.
\end{equation}
where $g$ is the determinant of metric tensor $g_{\mu\nu}$. Let us define
\begin{equation} \label{define3D}
    \left\{
        \begin{aligned}
            \boldsymbol{D}/\epsilon_0&=\sqrt{-g}(F^{01},F^{02},F^{03})\\
            Z_0\boldsymbol{H}&=\sqrt{-g}(F^{23},F^{31},F^{12})\\
            \boldsymbol{E}&=(F_{10},F_{20},F_{30})\\
            c\boldsymbol{B}&=(F_{23},F_{31},F_{12})\\
            c\rho&=\sqrt{-g}J^0\\
            \boldsymbol{J}&=\sqrt{-g}(J^1,J^2,J^3)\\
            c\tilde{\rho}&=\sqrt{-g}\tilde{J}^0\\
            \tilde{\boldsymbol{J}}&=\sqrt{-g}(\tilde{J}^1,\tilde{J}^2,\tilde{J}^3)
        \end{aligned}
    \right.
\end{equation}
where $\epsilon_0$ is vacuum permittivity, and $c$ is the vacuum light speed. By this definition, Eq.(\ref{Maxwell-partial derivative}) can be written as
\begin{equation}
    \left\{
        \begin{aligned}
            &\nabla\cdot\boldsymbol{D}/\epsilon_0=\rho/\epsilon_0\\
            &\frac{\partial}{\partial x^0}\boldsymbol{D}/\epsilon_0=\nabla\times Z_0\boldsymbol{H}-Z_0\boldsymbol{J}\\
            &\nabla\cdot c\boldsymbol{B}=c\tilde{\rho}\\
            &\frac{\partial c\boldsymbol{B}}{\partial x^0}=-\nabla\times\boldsymbol{E}-\tilde{\boldsymbol{J}}
        \end{aligned}
    \right.
\end{equation}
The constitutive equations are
\begin{equation}
    \left\{
        \begin{aligned}
            \boldsymbol{D}/\epsilon_0&=\bar{\boldsymbol{G}}^{de}\cdot\boldsymbol{E}+\bar{\boldsymbol{G}}^{db}\cdot c\boldsymbol{B}\\
            Z_0\boldsymbol{H}&=\bar{\boldsymbol{G}}^{he}\cdot\boldsymbol{E}+\bar{\boldsymbol{G}}^{hb}\cdot c\boldsymbol{B}
        \end{aligned}
    \right.
\end{equation}
where the matrices $\bar{\boldsymbol{G}}^{de}$, $\bar{\boldsymbol{G}}^{db}$, $\bar{\boldsymbol{G}}^{he}$, $\bar{\boldsymbol{G}}^{hb}$  can be deduced by raising tensor indices:
\begin{equation} \label{F=gF}
    F^{\mu\nu}=g^{\mu\alpha}g^{\nu\beta}F_{\alpha\beta}
\end{equation}

The line element in Schwarzschild space-time is
\begin{equation}
	\begin{aligned}
    ds^2=&-\left(1-\frac{R_S}{r}\right)(cdt)^2+\left(1-\frac{R_S}{r}\right)^{-1}dr^2\\
    	&+r^2(d\theta^2+\sin^2\theta d\phi^2),\quad r>R_S
	\end{aligned}
\end{equation}
where $R_S$ is Schwarzschild radius. The Cartesian coordinates are defined as
\begin{equation}
    x=R\sin\theta\cos\phi,\quad y=R\sin\theta\sin\phi,\quad z=R\cos\theta
\end{equation}
where \cite{20}
\begin{equation}
    R=\frac{1}{2}\left(r-\frac{R_S}{2}+\sqrt{r(r-R_S)}\right),\quad r>R_S
\end{equation}
The line element in the Cartesian coordinates is
\begin{equation}
    ds^2=-\left(\frac{1-\frac{R_S}{4R}}{1+\frac{R_S}{4R}}\right)^2(cdt)^2+\left(1+\frac{R_S}{4R}\right)^4(dx^2+dy^2+dz^2)
\end{equation}
The constitutive equations are
\begin{equation}
    \left\{
        \begin{aligned}
            \boldsymbol{D}&=\epsilon_0\left(1-\frac{R_S}{4R}\right)^{-1}\left(1+\frac{R_S}{4R}\right)^3\boldsymbol{E}\\
            \boldsymbol{B}&=\mu_0\left(1-\frac{R_S}{4R}\right)^{-1}\left(1+\frac{R_S}{4R}\right)^3\boldsymbol{H}
        \end{aligned}
    \right.
\end{equation}
where $\mu_0=Z_0/c$ is permeability of vacuum. The above equations can be deduced from Eq.(\ref{define3D}) and (\ref{F=gF}). The Schwarzschild space-time is equivalent to flat space-time which is filled with medium, and the relative permittivity and permeability of the medium are \cite{9}
\begin{equation}\label{er ur}
    \epsilon_r=\mu_r=\left(1-\frac{R_S}{4R}\right)^{-1}\left(1+\frac{R_S}{4R}\right)^3
\end{equation}
The FDTD method for inhomogeneous medium in Cartesian coordinate can be invoked directly in Schwarzschild space-time.

\section{Green function in curved space-time}\label{sec. green function}
	\subsection{Green function}
If $\tilde{J}=0$, we have $dF=0$ according to Eq.(\ref{Maxwell-exterior differential}). The electromagnetic tensor can be written as
\begin{equation}
	F_{\alpha\beta}=c(dA)_{\alpha\beta}=c(A_{\beta;\alpha}-A_{\alpha;\beta})
\end{equation}
where $A^\alpha$ is electric potential. Substituting the above equation into the first one in Eq.(\ref{Maxwell-covariant derivative}) and applying Lorentz gauge, we get the wave equation
\begin{equation}
	\Box A^\alpha-R^\alpha{}_\beta A^\beta=-\mu_0J^\alpha 
\end{equation}
where $\Box A^\alpha=A^\alpha{}_{;\beta}{}^{;\beta}$ and $R^\alpha{}_\beta$ is Ricci tensor. The electric field and magnetic field are
\begin{equation} \label{E_Je}
	\begin{aligned}
		E_i&=F_{i0}\\
		&=c(A_{0;1}-A_{1;0},A_{0;2}-A_{2;0},A_{0;3}-A_{3;0})
	\end{aligned}
\end{equation}
\begin{equation} \label{H_Je}
	\begin{aligned}
		Z_0H_i&=\sqrt{-g}\epsilon_{ijk}F^{jk}\\
		&=c\sqrt{-g}(A^{3;2}-A^{2;3},A^{1;3}-A^{3;1},A^{2;1}-A^{1;2})
	\end{aligned}
\end{equation} 
where the indices $i,j,k$ run from 1 to 3, and $\epsilon_{ijk}$ is Levi-Civita symbol.

Let's define the dual electromagnetic tensor $\tilde{F}$ as below
\begin{equation}
	\tilde{F}=*F\quad \text{or}\quad F=-*\tilde{F}
\end{equation}
The Eq.(\ref{Maxwell-exterior differential}) can be written as
\begin{equation} \label{Maxwell-exterior differential-dual}
    \left\{
        \begin{aligned}
            d*\tilde{F}&=-*\tilde{J}\\
            d\tilde{F}&=Z_0*J
        \end{aligned}
    \right.
\end{equation}
The Eq.(\ref{Maxwell-covariant derivative}) can be written as
\begin{equation} \label{Maxwell-covariant derivative-dual}
    \left\{
        \begin{aligned}
            &\tilde{F}^{\mu\nu}{}_{;\nu}=-\tilde{J}^\mu\\
            &\tilde{F}_{\nu\sigma;\mu}+\tilde{F}_{\mu\nu;\sigma}+\tilde{F}_{\sigma\mu;\nu}=Z_0(*J)_{\mu\nu\sigma}
        \end{aligned}
    \right.
\end{equation}
If $J=0$, we have $d\tilde{F}=0$ according to Eq.(\ref{Maxwell-exterior differential-dual}). The dual electromagnetic tensor can be written as
\begin{equation}
	\tilde{F}_{\alpha\beta}=-\frac{1}{\epsilon_0}(d\tilde{A})_{\alpha\beta}=-\frac{1}{\epsilon_0}(\tilde{A}_{\beta;\alpha}-\tilde{A}_{\alpha;\beta})
\end{equation}
where $\tilde{A}^\alpha$ is magnetic potential. Substituting the above equation into the first one in Eq.(\ref{Maxwell-covariant derivative-dual}) and applying Lorentz gauge, we get the wave equation
\begin{equation}
	\Box \tilde{A}^\alpha-R^\alpha{}_\beta\tilde{A}^\beta=-\epsilon_0\tilde{J}^\alpha 
\end{equation}
The electric field and magnetic field are
\begin{equation} \label{E_Jm}
	\begin{aligned}
		E_i&=-(*\tilde{F})_{i0}=\sqrt{-g}(\tilde{F}^{23},\tilde{F}^{31},\tilde{F}^{12})\\
		&=-\frac{\sqrt{-g}}{\epsilon_0}(\tilde{A}^{3;2}-\tilde{A}^{2;3},\tilde{A}^{1;3}-\tilde{A}^{3;1},\tilde{A}^{2;1}-\tilde{A}^{1;2})
	\end{aligned}
\end{equation}
\begin{equation} \label{H_Jm}
	\begin{aligned}
		Z_0H_i&=-\sqrt{-g}\epsilon_{ijk}(*\tilde{F})^{jk}=-(\tilde{F}_{10},\tilde{F}_{20},\tilde{F}_{30})\\
		&=\frac{1}{\epsilon_0}(\tilde{A}_{0;1}-\tilde{A}_{1;0},\tilde{A}_{0;2}-\tilde{A}_{2;0},\tilde{A}_{0;3}-\tilde{A}_{3;0})
	\end{aligned}
\end{equation} 
The total electric field is the summation of Eq.(\ref{E_Je}) and Eq.(\ref{E_Jm}), and the total magnetic field is the summation of Eq.(\ref{H_Je}) and Eq.(\ref{H_Jm}). We need to calculate covariant derivative of potential. 

The electric potential and magnetic potential can be expressed in integral form
\begin{equation}
	A_\alpha(x)=\frac{\mu_0}{4\pi}\int{G_{\alpha\beta'}(x,x')J^{\beta'}(x')\sqrt{-g(x')}d^4x'}
\end{equation}
\begin{equation}
	\tilde{A}_\alpha(x)=\frac{\epsilon_0}{4\pi}\int{G_{\alpha\beta'}(x,x')\tilde{J}^{\beta'}(x')\sqrt{-g(x')}d^4x'}
\end{equation}
where $G_{\alpha\beta'}$ is Green function. The Hadamard form of Green function is \cite{17}
\begin{equation}\label{Green}
\begin{aligned}
	&G_{\alpha\beta'}(x,x')=U_{\alpha\beta'}(x,x')\delta(\sigma(x,x'))\\
	&\quad+V_{\alpha\beta'}(x,x')\theta(-\sigma(x,x'))
\end{aligned}
\end{equation}
In the above expression, $U_{\alpha\beta'}$ and $V_{\alpha\beta'}$ are two bi-tensors, $\delta$ is Dirac function, and $\theta$ is the step function
\begin{equation}
	\theta(\sigma)=\left\{\begin{array}{l}
		1,\quad \sigma\geq 0\\
		0,\quad \sigma<0
	\end{array}\right.
\end{equation}
The bi-scalar $\sigma(x,x')$ in Eq.(\ref{Green}) is Synge's world function \cite{20} which is defined as half the square geodesic distance between the points $x$ and $x'$. For a specified $x'$, Green function has two branches - one is the chronological future and the other is the chronological past. Only the chronological future branch is taken into consideration in this paper. 

	\subsection{$U_{\alpha\beta'}$ and $V_{\alpha\beta'}$}
The bi-tensor $U_{\alpha\beta'}$ in Eq. (\ref{Green}) obeys transport equation \cite{19}
\begin{equation}\label{trans U_ab'}
	U_{\alpha\beta';\gamma}\sigma^\gamma+\frac{1}{2}(\sigma^\gamma{}_\gamma-4)U_{\alpha\beta'}=0
\end{equation}
The bi-tensors $\sigma^\alpha$ and $\sigma^\alpha{}_\beta$ are one and two order covariant derivative of $\sigma$. The bi-tensor $V_{\alpha\beta'}$ in Eq.(\ref{Green}) can be expended into power series of $\sigma$:
\begin{equation}
	V_{\alpha\beta'}(x,x')=\sum_{p=0}^\infty V^p_{\alpha\beta'}(x,x')[\sigma(x,x')]^p
\end{equation}
where the superscript $p$ in coefficients $V^p_{\alpha\beta'}$ is a serial number. $V^p_{\alpha\beta'}$ obey transport equations \cite{19}
\begin{equation}
	\left\lbrace\begin{aligned}
		&V^p_{\alpha\beta';\gamma}\sigma^\gamma+\left(\frac{1}{2}\sigma^\gamma{}_\gamma+p-1\right)V^p_{\alpha\beta'}\\
		&\quad+\frac{1}{2p}(\Box V^{p-1}_{\alpha\beta'}-R_\alpha{}^\mu V^{p-1}_{\mu\beta'})=0,\quad p>0\\
		&V^0_{\alpha\beta';\gamma}\sigma^\gamma+\frac{1}{2}(\sigma^\gamma{}_\gamma-2)V^0_{\alpha\beta'}=\frac{1}{2}(\Box U_{\alpha\beta'}-R_\alpha{}^\mu U_{\mu\beta'})
	\end{aligned}\right.	
\end{equation}
Numerical results show that, the part of $V_{\alpha\beta'}$ which is defined on time-like geodesic ($\sigma<0$) has little effect on Green function. Therefore, this part of $V_{\alpha\beta'}$ can be ignored and we have $V_{\alpha\beta'}\approx V^0_{\alpha\beta'}$. $U_{\alpha\beta'}$ and $V_{\alpha\beta'}$ can be solved from their transport equations \cite{18,19}.

The null geodesic ($\sigma=0$) that links $x$ to $x'$ is described by a relation function $z(t)$ in which $t$ is an affine parameter that ranges from 0 to 1. We have $z(0)=x'$ and $z(1)=x$. The function $z(t)$ is the solution of the boundary value problem
\begin{equation} \label{geodesic equation}
	\left\lbrace\begin{aligned}
		&\frac{d^2z^\alpha}{dt^2}+\varGamma^\alpha{}_{\mu\nu}\frac{dz^\mu}{dt}\frac{dz^\nu}{dt}=0\\
		&z(0)=x'\\
		&z^i(1)=x^i, i=1,2,3\\
		&g_{\mu\nu}(x')u^\mu(0)u^\nu(0)=0
	\end{aligned}\right.		
\end{equation}
where $\varGamma^\alpha{}_{\mu\nu}$ is Christoffel symbol, $u^\mu=dz^\mu/dt$. This boundary value problem can be solved by Lobatto IIIa formula. For a specified $x'$, $U_{\alpha\beta'}(z,x')$ is a function of $t$, and it can be denoted as $U_{\alpha\beta'}(t)$. Thus, Eq.(\ref{trans U_ab'}) can be rewritten as an initial value problem
\begin{equation}\label{U_ab'}
	\left\lbrace\begin{aligned}
		&\frac{dU_{\alpha\beta'}}{dt}-\varGamma^\rho{}_{\alpha\mu}U_{\rho\beta'}u^\mu+\frac{1}{2t}(\sigma^\mu{}_\mu-4)U_{\alpha\beta'}=0\\
		&U_{\alpha\beta'}(0)=g_{\alpha'\beta'}
	\end{aligned}\right.		
\end{equation}
where we have applied relation $\sigma^\alpha=tu^\alpha$ and the coincidence limit [17] of $U_{\alpha\beta'}$. The initial value problem can be solved by Runge-Kutta formula. After $U_{\alpha\beta'}(t)$ is solved from Eq.(\ref{U_ab'}), $U_{\alpha\beta'}(1)$ is exactly the $U_{\alpha\beta'}(x,x')$ in Eq.(\ref{Green}). In the same way, $V^0_{\alpha\beta'}$ can be solved from
\begin{equation}\label{V_ab'}
	\left\lbrace\begin{aligned}
		&\frac{dV^0_{\alpha\beta'}}{dt}-\varGamma^\rho{}_{\alpha\mu}V^0_{\rho\beta'}u^\mu+\frac{1}{2t}(\sigma^\mu{}_\mu-4)V^0_{\alpha\beta'}\\
		&\quad+\frac{1}{t}\left[V^0_{\alpha\beta'}-\frac{1}{2}(\Box U_{\alpha\beta'}-R_\alpha{}^\mu U_{\mu\beta'})\right]=0\\
		&V^0_{\alpha\beta'}(0)=\frac{1}{12}g_{\alpha'\beta'}R'
	\end{aligned}\right.
\end{equation}
where $R'$ is scalar curvature at $x'$. $\sigma^\alpha{}_\beta$ and $\Box U_{\alpha\beta'}$ are unknown quantities which have to be solved before solving Eq.(\ref{U_ab'}) and (\ref{V_ab'}).

	\subsection{$\sigma^\alpha{}_\beta$ and $\Box U_{\alpha\beta'}$}
Differentiating $\sigma^\alpha\sigma_\alpha=2\sigma$ twice, we obtain
\begin{equation}
	\sigma^\alpha{}_{\mu\beta}\sigma^\mu+\sigma^\alpha{}_\mu\sigma^\mu{}_\beta=\sigma^\alpha{}_\beta
\end{equation}
By commuting covariant derivatives, i.e., substituting $\sigma^\alpha{}_{\mu\beta}=\sigma^\alpha{}_{\beta\mu}-R^\alpha{}_{\rho\mu\beta}\sigma^\rho$ ($R^\alpha{}_{\beta\gamma\delta}$ is Riemann curvature tensor) into the above equation we obtain
\begin{equation}\label{trans Sa_b}
	\sigma^\alpha{}_{\beta\mu}\sigma^\mu-R^\alpha{}_{\rho\mu\beta}\sigma^\rho\sigma^\mu+(\sigma^\alpha{}_\mu-\delta^\alpha{}_\mu)\sigma^\mu{}_\beta=0
\end{equation}
The above equation can be rewritten as an initial value problem
\begin{equation}\label{Sa_b}
	\left\lbrace\begin{aligned}
		&\frac{d\sigma^\alpha{}_\beta}{dt}=(\sigma^\alpha{}_\rho\varGamma^\rho{}_{\mu\beta}-\varGamma^\alpha{}_{\rho\mu}\sigma^\rho{}_\beta)u^\mu+tR^\alpha{}_{\rho\mu\beta}u^\rho u^\mu\\
		&\quad-\frac{1}{t}(\sigma^\alpha{}_\mu-\delta^\alpha{}_\mu)\sigma^\mu{}_\beta\\
		&\sigma^\alpha{}_\beta(0)=\delta^{\alpha'}{}_{\beta'}
	\end{aligned}\right.
\end{equation}
$\sigma^\alpha{}_\beta$ can be solved from the above equation.

By differentiating Eq.(\ref{trans U_ab'}) and commuting covariant derivatives, we obtain
\begin{equation}\label{trans U_ab'c}
	\begin{aligned}
	&U_{\alpha\beta';\gamma\mu}\sigma^\mu+R^\rho{}_{\alpha\mu\gamma}U_{\rho\beta'}\sigma^\mu+U_{\alpha\beta';\mu}\sigma^\mu{}_\gamma+\frac{1}{2}U_{\alpha\beta'}\sigma^\mu{}_{\mu\gamma}\\
	&\quad+\frac{1}{2}(\sigma^\mu{}_\mu-4)U_{\alpha\beta';\gamma}=0
	\end{aligned}
\end{equation}
It can be rewritten as an initial value problem
\begin{equation}\label{U_ab'c}
	\left\lbrace\begin{aligned}
		&\frac{dU_{\alpha\beta';\gamma}}{dt}-(\varGamma^\rho{}_{\alpha\mu}U_{\rho\beta';\gamma}+\varGamma^\rho{}_{\gamma\mu}U_{\alpha\beta';\rho})u^\mu\\
		&\quad+\frac{1}{t}U_{\alpha\beta';\mu}\sigma^\mu{}_\gamma+\frac{1}{2t}U_{\alpha\beta'}\sigma^\mu{}_{\mu\gamma}\\
		&\quad+\frac{1}{2t}(\sigma^\mu{}_\mu-4)U_{\alpha\beta';\gamma}=0\\
		&U_{\alpha\beta';\gamma}(0)=0
	\end{aligned}\right.
\end{equation}
$U_{\alpha\beta';\gamma}$ can be solved from the above equation. By differentiating Eq.(\ref{trans U_ab'c}) and commuting covariant derivatives, an initial value problem can be obtained:
\begin{equation}\label{U_ab'cd}
	\left\{\begin{aligned}
		&\frac{dU_{\alpha\beta';\gamma\delta}}{dt}-(\varGamma^\rho{}_{\alpha\mu}U_{\rho\beta';\gamma\delta}+\varGamma^\rho{}_{\gamma\mu}U_{\alpha\beta';\rho\delta}\\
		&\quad+\varGamma^\rho{}_{\delta\mu}U_{\rho\beta';\gamma\rho})u^\mu\\
		&\quad+(R^\rho{}_{\alpha\mu\gamma;\delta}U_{\rho\beta'}+R^\rho{}_{\alpha\mu\gamma}U_{\rho\beta';\delta}+R^\rho{}_{\alpha\mu\delta}U_{\rho\beta';\gamma}\\
		&\quad+R^\rho{}_{\gamma\mu\delta}U_{\alpha\beta';\rho})u^\mu\\
		&\quad+\frac{1}{t}\left(U_{\alpha\beta';\mu\gamma}\sigma^\mu{}_\delta+U_{\alpha\beta';\mu\delta}\sigma^\mu{}_\gamma+\frac{1}{2}U_{\alpha\beta'}\sigma^\mu{}_{\mu\gamma\delta}\right)\\
		&\quad+\frac{1}{t}U_{\alpha\beta';\mu}\sigma^\mu{}_{\gamma\delta}+\frac{1}{2t}U_{\alpha\beta';\delta}\sigma^\mu{}_{\mu\gamma}+\frac{1}{2t}U_{\alpha\beta';\gamma}\sigma^\mu{}_{\mu\delta}\\
		&\quad+\frac{1}{2t}(\sigma^\mu{}_\mu-4)U_{\alpha\beta';\gamma\delta}=0\\
		&U_{\alpha\beta';\gamma\delta}(0)=-\frac{1}{2}R_{\alpha'\beta'\gamma'\delta'}+\frac{1}{6}g_{\alpha'\beta'}R_{\gamma'\delta'}
	\end{aligned}\right.
\end{equation}
$U_{\alpha\beta';\gamma\delta}$ can be solved from the above initial value problem and $\Box U_{\alpha\beta'}$ can be obtained from
\begin{equation}\label{Box U}
	\Box U_{\alpha\beta'}=g^{\mu\nu}U_{\alpha\beta';\mu\nu}
\end{equation}
In Eqs.(\ref{U_ab'c})(\ref{U_ab'cd}), $\sigma^\alpha{}_{\beta\gamma}$ and $\sigma^\alpha{}_{\beta\gamma\delta}$ are still unknown quantities.

	\subsection{$\sigma^\alpha{}_{\beta\gamma}$ and $\sigma^\alpha{}_{\beta\gamma\delta}$}
Through differentiating Eq.(\ref{trans Sa_b}) and commuting covariant derivatives, the following equation can be obtained
\begin{equation}\label{trans Sa_bc}
	\begin{aligned}
		&\sigma^\alpha{}_{\beta\gamma\mu}\sigma^\mu+\sigma^\alpha{}_{\mu\beta}\sigma^\mu{}_\gamma+\sigma^\alpha{}_{\mu\gamma}\sigma^\mu{}_\beta+(\sigma^\alpha{}_\mu-\delta^\alpha{}_\mu)\sigma^\mu{}_{\beta\gamma}\\
		&\quad-R^\alpha{}_{\rho\mu\beta;\gamma}\sigma^\rho\sigma^\mu-R^\alpha{}_{\rho\mu\beta}\sigma^\rho{}_\gamma\sigma^\mu-R^\alpha{}_{\rho\mu\gamma}\sigma^\rho{}_\beta\sigma^\mu\\
		&\quad+R^\rho{}_{\beta\mu\gamma}\sigma^\alpha{}_\rho\sigma^\mu=0
	\end{aligned}
\end{equation}
It can be rewritten as an initial value problem
\begin{equation}\label{Sa_bc}
	\left\{\begin{aligned}
		&\frac{d\sigma^\alpha{}_{\beta\gamma}}{dt}+(\varGamma^\alpha{}_{\rho\mu}\sigma^\rho{}_{\beta\gamma}-\varGamma^\rho{}_{\beta\mu}\sigma^\alpha{}_{\rho\gamma}-\varGamma^\rho{}_{\gamma\mu}\sigma^\alpha{}_{\beta\rho})u^\mu\\
		&\quad+\frac{1}{t}\sigma^\alpha{}_{\mu\beta}\sigma^\mu{}_\gamma+\frac{1}{t}\sigma^\alpha{}_{\mu\gamma}\sigma^\mu{}_\beta+\frac{1}{t}(\sigma^\alpha{}_\mu-\delta^\alpha{}_\mu)\sigma^\mu{}_{\beta\gamma}\\
		&\quad-(R^\alpha{}_{\rho\mu\beta;\gamma}\sigma^\rho+R^\alpha{}_{\rho\mu\beta}\sigma^\rho{}_\gamma+R^\alpha{}_{\rho\mu\gamma}\sigma^\rho{}_\beta\\
		&\quad-R^\alpha{}_{\beta\mu\gamma}\sigma^\alpha{}_\rho)u^\mu=0\\
		&\sigma^\alpha{}_{\beta\gamma}(0)=0
	\end{aligned}\right.
\end{equation}
$\sigma^\alpha{}_{\beta\gamma}$ can be solved from the above initial value problem. Through differentiating Eq.(\ref{trans Sa_bc}) and commuting covariant derivatives, an initial value problem can be obtained, from which $\sigma^\alpha{}_{\beta\gamma\delta}$ can be solved:
\begin{equation}\label{Sa_bcd}
	\left\{\begin{aligned}
		&\frac{d\sigma^\alpha{}_{\beta\gamma\delta}}{dt}+(\varGamma^\alpha{}_{\rho\mu}\sigma^\rho{}_{\beta\gamma\delta}-\varGamma^\rho{}_{\beta\mu}\sigma^\alpha{}_{\rho\gamma\delta}-\varGamma^\rho{}_{\gamma\mu}\sigma^\alpha{}_{\beta\rho\delta}\\
		&\quad-\varGamma^\rho{}_{\delta\mu}\sigma^\alpha{}_{\beta\gamma\rho})u^\mu\\
		&\quad-(R^\alpha{}_{\rho\mu\beta;\gamma}\sigma^\rho{}_\delta+R^\alpha{}_{\rho\mu\beta;\delta}\sigma^\rho{}_\gamma+R^\alpha{}_{\rho\mu\gamma;\delta}\sigma^\rho{}_\beta\\
		&\quad-R^\rho{}_{\beta\mu\delta}\sigma^\alpha{}_{\rho\gamma}-R^\rho{}_{\beta\mu\gamma}\sigma^\alpha{}_{\rho\delta}-R^\rho{}_{\gamma\mu\delta}\sigma^\alpha{}_{\beta\rho}\\
		&\quad+R^\alpha{}_{\rho\mu\gamma}\sigma^\rho{}_{\beta\delta}+R^\alpha{}_{\rho\mu\delta}\sigma^\rho{}_{\beta\gamma}+R^\alpha{}_{\rho\mu\beta}\sigma^\rho{}_{\gamma\delta}\\
		&\quad-R^\rho{}_{\beta\mu\gamma;\delta}\sigma^\alpha{}_\rho)u^\mu-tR^\alpha{}_{\rho\mu\beta;\gamma\delta}u^\rho u^\mu\\
		&\quad+\frac{1}{t}(\sigma^\alpha{}_{\mu\beta\gamma}\sigma^\mu{}_\delta+\sigma^\alpha{}_{\mu\gamma\delta}\sigma^\mu{}_\beta+\sigma^\alpha{}_{\mu\beta\delta}\sigma^\mu{}_\gamma)\\
		&\quad+\frac{1}{t}\sigma^\alpha{}_{\mu\beta}\sigma^\mu{}_{\gamma\delta}+\frac{1}{t}\sigma^\alpha{}_{\mu\gamma}\sigma^\mu{}_{\beta\delta}+\frac{1}{t}\sigma^\alpha{}_{\mu\delta}\sigma^\mu{}_{\beta\gamma}\\
		&\quad+\frac{1}{t}(\sigma^\alpha{}_\mu-\delta^\alpha{}_\mu)\sigma^\mu{}_{\beta\gamma\delta}=0\\
		&\sigma^\alpha{}_{\beta\gamma\delta}(0)=-\frac{1}{3}(R^{\alpha'}{}_{\gamma'\beta'\delta'}+R^{\alpha'}{}_{\delta'\beta'\gamma'})
	\end{aligned}\right.
\end{equation}

	 \subsection{Limit formulas}
In the above initial value problems, i.e., Eqs.(\ref{U_ab'})(\ref{V_ab'})(\ref{Sa_b})(\ref{U_ab'c})(\ref{U_ab'cd})(\ref{Sa_bc})(\ref{Sa_bcd}), there are terms involving $1/t$. Evaluating these terms is indispensable when solving differential equations. However, to evaluate them directly at $t=0$ is impossible. Fortunately, the terms have limit formulas which can be derived from their covariant expansions \cite{17,19}.

The covariant expansion of $\sigma^\alpha{}_\beta$ is
\begin{equation}\label{expension Sa_b}
	\sigma^\alpha{}_\beta=\delta^\alpha{}_\beta-\frac{1}{3}g^{\alpha\alpha'}g^{\beta'}{}_\beta R_{\alpha'\mu'_1\beta'\mu'_2}t^2u^{\mu'_1}u^{\mu'_2}+O(t^3)
\end{equation}
where $g^\alpha{}_{\beta'}$ is parallel propagator. By moving the term $\delta^\alpha{}_\beta$ to the left side, dividing it by $t$, and then taking limit, the following equation can be set up
\begin{equation}\label{lim Sa_b}
	\left(\frac{\sigma^\alpha{}_\beta-\delta^\alpha{}_\beta}{t}\right)_{t\to 0}=0
\end{equation}
By taking the contraction, we get
\begin{equation}
	\left(\frac{\sigma^\alpha{}_\alpha-4}{t}\right)_{t\to 0}=0
\end{equation}
The covariant expansion of $\sigma_{\alpha\beta\gamma}$ is
\begin{equation}
\begin{aligned}
	&\sigma_{\alpha\beta\gamma}=g^{\alpha'}{}_\alpha g^{\beta'}{}_\beta g^{\gamma'}{}_\gamma\left[-\frac{1}{3}(R_{\alpha'\gamma'\beta'\mu'}+R_{\alpha'\mu'\beta'\gamma'})tu^{\mu'}\right.\\
	&\left.\quad+O(t^2)\right]
\end{aligned}
\end{equation}
from which we get
\begin{equation}\label{lim S_abc}
	\left(\frac{\sigma_{\alpha\beta\gamma}}{t}\right)_{t\to 0}=-\frac{1}{3}(R_{\alpha'\gamma'\beta'\mu'}+R_{\alpha'\mu'\beta'\gamma'})u^{\mu'}
\end{equation}
By taking the contraction, we get
\begin{equation}
	\left(\frac{\sigma^{\mu}{}_{\mu\gamma}}{t}\right)_{t\to 0}=-\frac{2}{3}R_{\gamma'\mu'}u^{\mu'}
\end{equation}
The covariant expansion of $\sigma_{\alpha\beta\gamma\delta}$ is
\begin{equation}\label{expansion Sa_bcd}
\begin{aligned}
	&\sigma_{\alpha\beta\gamma\delta}=g^{\alpha'}{}_\alpha g^{\beta'}{}_\beta g^{\gamma'}{}_\gamma g^{\delta'}{}_\delta\left\{-\frac{1}{3}(R_{\alpha'\gamma'\beta'\delta'}+R_{\alpha'\delta'\beta'\gamma'})\right.\\
	&\left.\quad+[\sigma_{\alpha\beta\gamma\delta\mu}]tu^{\mu'}+O(t^2)\right\}
\end{aligned}
\end{equation}
where the coincidence limit is
\begin{equation}
	\begin{aligned}
		&[\sigma_{\alpha_1\alpha_2\alpha_3\alpha_4\alpha_5}]=-\frac{1}{4}(R_{\alpha'_5\alpha'_1\alpha'_3\alpha'_2;\alpha'_4}+R_{\alpha'_4\alpha'_1\alpha'_3\alpha'_2;\alpha'_5}\\
		&\quad+R_{\alpha'_5\alpha'_1\alpha'_4\alpha'_2;\alpha'_3}+R_{\alpha'_3\alpha'_1\alpha'_4\alpha'_2;\alpha'_5}+R_{\alpha'_4\alpha'_1\alpha'_5\alpha'_2;\alpha'_3}\\
		&\quad+R_{\alpha'_3\alpha'_1\alpha'_5\alpha'_2;\alpha'_4})
	\end{aligned}
\end{equation}
which can be derived by differentiating $\sigma^\alpha\sigma_\alpha=2\sigma$ for five times, and then taking coincidence limit. From Eq.( \ref{expension Sa_b}) and (\ref{expansion Sa_bcd}) we get
\begin{equation}
	\begin{aligned}
		&\left[\frac{1}{t}(\sigma_{\alpha\mu\gamma\delta}\sigma^\mu{}_\beta+\sigma_{\alpha\mu\beta\delta}\sigma^\mu{}_\gamma+\sigma_{\alpha\mu\beta\gamma}\sigma^\mu{}_\delta)\right]_{t\to 0}=\\
		&\quad\frac{1}{4}(R_{\alpha'\beta'\mu'\gamma';\delta'}+R_{\alpha'\beta'\mu'\delta';\gamma'}+R_{\alpha'\gamma'\mu'\beta';\delta'}\\
		&\quad+R_{\alpha'\gamma'\mu'\delta';\beta'}+R_{\alpha'\delta'\mu'\beta';\gamma'}+R_{\alpha'\delta'\mu'\gamma';\beta'})u^{\mu'}
	\end{aligned}
\end{equation}
The covariant expansion of $U_{\alpha\beta';\gamma}$ is
\begin{equation}
\begin{aligned}
	&U_{\alpha\beta';\gamma}=g^{\alpha'}{}_\alpha g^{\gamma'}{}_\gamma\left[\left(-\frac{1}{2}R_{\alpha'\beta'\gamma'\mu'}+\frac{1}{6}g_{\alpha'\beta'}R_{\gamma'\mu'}\right)tu^{\mu'}\right.\\
	&\left.\quad+O(t^2)\right]
\end{aligned}
\end{equation}
from which we get
\begin{equation}\label{lim U_ab'c}
	\left(\frac{U_{\alpha\beta';\gamma}}{t}\right)_{t\to 0}=\left(-\frac{1}{2}R_{\alpha'\beta'\gamma'\mu'}+\frac{1}{6}g_{\alpha'\beta'}R_{\gamma'\mu'}\right)u^{\mu'}
\end{equation}
The covariant expansion of $U_{\alpha\beta';\gamma\delta}$ is
\begin{equation}\label{expansion U_ab'cd}
	\begin{aligned}
		&U_{\alpha\beta';\alpha_1\alpha_2}=g^{\alpha'}{}_\alpha g^{\alpha'_1}{}_{\alpha_1}g^{\alpha'_2}{}_{\alpha_2}\left\{-\frac{1}{2}\left(R_{\alpha'\beta'\alpha'_1\alpha'_2}\right.\right.\\
		&\quad\left.-\frac{1}{3}g_{\alpha'\beta'}R_{\alpha'_1\alpha'_2}\right)-\frac{1}{3}\left[R_{\alpha'\beta'\alpha'_1\alpha'_2;\mu'}\right.\\
		&\quad+R_{\alpha'\beta'\alpha'_1\mu';\alpha'_2}-\frac{1}{4}g_{\alpha'\beta'}(R_{\alpha'_1\alpha'_2;\mu'}+R_{\alpha'_2\mu';\alpha'_1}\\
		&\quad\left.\left.+R_{\alpha'_1\mu';\alpha'_2})\right]tu^{\mu'}+O(t^2)\right\}
	\end{aligned}
\end{equation}
From Eqs.(\ref{expension Sa_b})(\ref{expansion Sa_bcd})(\ref{expansion U_ab'cd}) we get
\begin{equation}
	\begin{aligned}	
		&\left[\frac{1}{t}\left(U_{\alpha\beta';\mu\gamma}\sigma^\mu{}_\delta+U_{\alpha\beta';\mu\delta}\sigma^\mu{}_\gamma+\frac{1}{2}U_{\alpha\beta'}\sigma^\mu{}_{\mu\gamma\delta}\right)\right]_{t\to 0}\\
		&\quad=-\frac{1}{3}\left[R_{\alpha'\beta'\gamma'\mu';\delta'}+R_{\alpha'\beta'\delta'\mu';\gamma'}+\frac{1}{4}g_{\alpha'\beta'}(R_{\gamma'\delta';\mu'}\right.\\
		&\quad\left.+R_{\delta'\mu';\gamma'}+R_{\gamma'\mu';\delta'})\right]u^{\mu'}
	\end{aligned}
\end{equation}
By taking the contraction of Eq.(\ref{expansion U_ab'cd}), we get
\begin{equation}\label{expansion Box U}
\begin{aligned}	
	&\Box U_{\alpha\beta'}=g^{\alpha'}{}_\alpha\left[\frac{1}{6}g_{\alpha'\beta'}R'-\frac{1}{3}\left(R^{\alpha'_1}{}_{\mu'\alpha'\beta';\alpha'_1}\right.\right.\\
	&\quad\left.\left.-\frac{1}{2}g_{\alpha'\beta'}R_{;\mu'}\right)tu^{\mu'}+O(t^2)\right]
\end{aligned}
\end{equation}
The covariant expansion of $V^0_{\alpha\beta'}$ is
\begin{equation}\label{expansion V0}
	\begin{aligned}	
		&V^0_{\alpha\beta'}=g^{\alpha'}{}_\alpha\left[\frac{1}{12}g_{\alpha'\beta'}R'-\frac{1}{2}R_{\alpha'\beta'}-\frac{1}{12}\left(R^{\gamma'}{}_{\mu'\alpha'\beta';\gamma'}\right.\right.\\
		&\quad\left.\left.-\frac{1}{2}g_{\alpha'\beta'}R_{;\mu'}\right)tu^{\mu'}-\frac{1}{4}R_{\alpha'\beta';\mu'}tu^{\mu'}+O(t^2)\right]
	\end{aligned}
\end{equation}
From Eq.(\ref{expansion Box U}) and (\ref{expansion V0}) we get
\begin{equation}
	\begin{aligned}	
		&\left\{\frac{1}{t}\left[V^0_{\alpha\beta'}-\frac{1}{2}(\Box U_{\alpha\beta'}-R_\alpha{}^\mu U_{\mu\beta'})\right]\right\}_{t\to 0}=\\
		&\quad\frac{1}{12}\left(R^{\gamma'}{}_{\mu'\alpha'\beta';\gamma'}-\frac{1}{2}g_{\alpha'\beta'}R_{;\mu'}+3R_{\alpha'\beta';\mu'}\right)u^{\mu'}
	\end{aligned}
\end{equation}

	 \subsection{The covariant derivative of potential}
The covariant derivative of potential is
\begin{equation}
	\begin{aligned}	
		&A_{\alpha;\gamma}=\frac{\mu_0}{4\pi}\int\left[U_{\alpha\beta'}\sigma_\gamma\delta'(\sigma)+(U_{\alpha\beta';\gamma}-V_{\alpha\beta'}\sigma_\gamma)\delta(\sigma)\right.\\
		&\quad\left.+V_{\alpha\beta';\gamma}\theta(-\sigma)\right]J^{\beta'}\sqrt{-g'}d^4x'
	\end{aligned}
\end{equation}
If ignoring $V^p_{\alpha\beta'}$ ($p>0$), the above formula can be written as
\begin{equation}\label{A_ab 0}
\begin{aligned}	
	&A_{\alpha;\gamma}=\frac{\mu_0}{4\pi}\int\left[U_{\alpha\beta'}\sigma_\gamma\delta'(\sigma)+(U_{\alpha\beta';\gamma}-V_{\alpha\beta'}\sigma_\gamma)\delta(\sigma)\right]\\
	&\quad J^{\beta'}\sqrt{-g'}d^4x'
\end{aligned}
\end{equation}
To calculate the above integral, we have to use the integral formula involving Dirac function:
\begin{equation}\label{int delta}
	\int \phi(t)\delta(f(t))dt=\sum_i\frac{\phi(t_i)}{|f'(t_i)|}
\end{equation}
\begin{equation}\label{int delta'}
	\int \phi(t)\delta'(f(t))dt=-\sum_i\frac{1}{|f'(t_i)|}\frac{d}{dt}\left(\frac{\phi(t)}{f'(t)}\right)_{t=t_i}
\end{equation}
where $f(t)$ and $\phi(t)$ are two arbitrary smooth functions and $t_i$ is the zero point of $f(t)$, i.e., $f(t_i)=0$. Applying Eqs.(\ref{int delta})(\ref{int delta'}) to Eq.(\ref{A_ab 0}), we obtain
\begin{equation}\label{A_ab}
	\begin{aligned}	
		&A_{\alpha;\gamma}=\frac{\mu_0}{4\pi}\int\left\{-\frac{\sqrt{-g'}}{|\sigma_{0'}|^3}\left[(U_{\alpha\beta',0'}J^{\beta'}+U_{\alpha\beta'}J^{\beta'}{}_{,0'})\sigma_\gamma\sigma_{0'}\right.\right.\\
		&\quad\left.+U_{\alpha\beta'}J^{\beta'}(\sigma_{\gamma,0'}\sigma_{0'}-\sigma_\gamma\sigma_{0',0'})\right]\\
		&\quad\left.+\frac{\sqrt{-g'}}{|\sigma_{0'}|}(U_{\alpha\beta';\gamma}-V_{\alpha\beta'}\sigma_\gamma)J^{\beta'}\right\}_{x'^0_z}d^3x'
	\end{aligned}
\end{equation}
where $x'^0_z$ is the zero point of $\sigma$ for fixed $x$ and $x'^i$ ($i=1,2,3$). Replacing $\mu_0$ and $J$ with $\epsilon_0$ and $\tilde{J}$ respectively, we can get the expression of $\tilde{A}_{\alpha;\gamma}$.

There are three unknown quantities $\sigma_{\alpha'\beta'}$, $\sigma_{\alpha\beta'}$ and $U_{\alpha\beta';\gamma'}$ on the right side of Eq.(\ref{A_ab}). We can get the transport equation of $\sigma^{\alpha'}{}_{\beta'}$ by substituting unprimed indices in Eq.(\ref{trans Sa_b}) into primed indices. Therefore, $\sigma^{\alpha'}{}_{\beta'}$ is the solution of the following initial problem which is similar to Eq.(\ref{Sa_b}):
\begin{equation}\label{Sa'_b'}
	\left\{\begin{aligned}	
		&\frac{d\sigma^{\alpha'}{}_{\beta'}}{dt}=(\sigma^{\alpha'}{}_{\rho'}\varGamma^{\rho'}{}_{\mu'\beta'}-\varGamma^{\alpha'}{}_{\rho'\mu'}\sigma^{\rho'}{}_{\beta'})\\
		&\qquad+tR^{\alpha'}{}_{\rho'\mu'\beta'}u^{\rho'}u^{\mu'}-\frac{1}{t}(\sigma^{\alpha'}{}_{\mu'}-\delta^{\alpha'}{}_{\mu'})\sigma^{\mu'}{}_{\beta'}\\
		&\sigma^{\alpha'}{}_{\beta'}(0)=\delta^\alpha{}_\beta
	\end{aligned}\right.
\end{equation}
Symmetric property holds for $\sigma_{\alpha\beta'}$, i.e., $\sigma_{\alpha\beta'}=\sigma_{\beta'\alpha}$. The transport equation of $\sigma_{\alpha'\beta}$ can be deduced by differentiating $\sigma^\mu\sigma_\mu=2\sigma$ at $x'$, and then differentiating it at $x$: 
\begin{equation}
	\sigma_{\alpha'\beta\mu}\sigma^\mu+\sigma_{\alpha'\mu}(\sigma^\mu{}_\beta-\delta^\mu{}_\beta)=0
\end{equation}
It can be rewritten as an initial value problem
\begin{equation}\label{S_a'b}
	\left\{\begin{aligned}	
		&\frac{d\sigma_{\alpha'\beta}}{dt}-\varGamma^{\rho}{}_{\beta\mu}\sigma_{\alpha'\rho}u^\mu+\frac{1}{t}\sigma_{\alpha'\mu}(\sigma^\mu{}_\beta-\delta^\mu{}_\beta)=0\\
		&\sigma_{\alpha'\beta}(0)=-g_{\alpha'\beta'}
	\end{aligned}\right.
\end{equation}
We can solve $\sigma_{\alpha'\beta}$ from the above equation. The transport equation of $U_{\alpha\beta';\gamma'}$ can be deduced by exchanging $x$ and $x'$ in Eq.(\ref{trans U_ab'}), and then differentiating it at $x'$. $U_{\alpha\beta';\gamma'}$ is the solution of the following initial problem:
\begin{equation}\label{U_ab'c'}
	\left\{\begin{aligned}	
		&\frac{dU_{\alpha\beta';\gamma'}}{dt}-(\varGamma^{\rho'}{}_{\beta'\mu'}U_{\alpha\rho';\gamma'}+\varGamma^{\rho'}{}_{\gamma'\mu'}U_{\alpha\beta';\rho'}\\
		&\quad+R^{\rho'}{}_{\beta'\gamma'\mu'}U_{\alpha'\rho'})u^{\mu'}+\frac{1}{t}U_{\alpha\beta';\mu'}\sigma^{\mu'}{}_{\gamma'}\\
		&\quad+\frac{1}{2t}U_{\alpha\beta'}\sigma^{\mu'}{}_{\mu'\gamma'}+\frac{1}{2t}(\sigma^{\mu'}{}_{\mu'}-4)U_{\alpha\beta';\gamma'}=0\\
		&U_{\alpha\beta';\gamma'}(0)=0
	\end{aligned}\right.
\end{equation}
The unknown quantity $\sigma^{\alpha'}{}_{\beta'\gamma'}$ in above equation is the solution of the following initial problem which is similar to Eq.(\ref{Sa_bc}):
\begin{equation}\label{Sa'_b'c'}
	\left\{\begin{aligned}	
		&\frac{d\sigma^{\alpha'}{}_{\beta'\gamma'}}{dt}+(\varGamma^{\alpha'}{}_{\rho'\mu'}\sigma^{\rho'}{}_{\beta'\gamma'}-\varGamma^{\rho'}{}_{\beta'\mu'}\sigma^{\alpha'}{}_{\rho'\gamma'}\\
		&\quad-\varGamma^{\rho'}{}_{\gamma'\mu'}\sigma^{\alpha'}{}_{\beta'\rho'}-R^{\alpha'}{}_{\rho'\mu'\beta';\gamma'}\sigma^{\rho'}-R^{\alpha'}{}_{\rho'\mu'\beta'}\sigma^{\rho'}{}_{\gamma'}\\
		&\quad-R^{\alpha'}{}_{\rho'\mu'\gamma'}\sigma^{\rho'}{}_{\beta'}+R^{\rho'}{}_{\beta'\mu'\gamma'}\sigma^{\alpha'}{}_{\rho'})u^{\mu'}\\
		&\quad+\frac{1}{t}\sigma^{\alpha'}{}_{\mu'\beta'}\sigma^{\mu'}{}_{\gamma'}+\frac{1}{t}\sigma^{\alpha'}{}_{\mu'\gamma'}\sigma^{\mu'}{}_{\beta'}\\
		&\quad+\frac{1}{t}(\sigma^{\alpha'}{}_{\mu'}-\delta^{\alpha'}{}_{\mu'})\sigma^{\mu'}{}_{\beta'\gamma'}=0\\		
		&\sigma^{\alpha'}{}_{\beta'\gamma'}(0)=0
	\end{aligned}\right.
\end{equation}
Limit formulas of terms involving $1/t$ in Eqs.(\ref{Sa'_b'})(\ref{S_a'b})(\ref{U_ab'c'})(\ref{Sa'_b'c'}) are similar to Eqs.(\ref{lim Sa_b})(\ref{lim S_abc})(\ref{lim U_ab'c}):
\begin{equation}
	\left(\frac{\sigma^{\alpha'}{}_{\beta'}-\delta^{\alpha'}{}_{\beta'}}{t}\right)_{t\to 0}=0
\end{equation}
\begin{equation}
	\left(\frac{\sigma_{\alpha'\beta'\gamma'}}{t}\right)_{t\to 0}=-\frac{1}{3}(R_{\alpha\gamma\beta\mu}+R_{\alpha\mu\beta\gamma})u^\mu
\end{equation}
\begin{equation}
	\left(\frac{U_{\alpha\beta';\gamma'}}{t}\right)_{t\to 0}=\left(\frac{1}{2}R_{\alpha\beta\gamma\mu}+\frac{1}{6}g_{\alpha\beta}R_{\gamma\mu}\right)u^\mu
\end{equation}

As a summary of this section, the solving sequence for the covariant derivative of potential is listed as below:

Eq.(\ref{geodesic equation}) $\to$ Eq.(\ref{Sa_b}) $\to$ Eq.(\ref{Sa_bc}) $\to$ Eq.(\ref{Sa_bcd}) $\to$ Eq.(\ref{Sa'_b'}) $\to$ Eq.(\ref{S_a'b}) $\to$ Eq.(\ref{Sa'_b'c'}) $\to$ Eq.(\ref{U_ab'}) $\to$ Eq.(\ref{U_ab'c}) $\to$ Eq.(\ref{U_ab'cd}) $\to$ Eq.(\ref{Box U}) $\to$ Eq.(\ref{V_ab'}) $\to$ Eq.(\ref{U_ab'c'}) $\to$ Eq.(\ref{A_ab})

\section{Connection boundary and output boundary}\label{sec. cb ob}
Electric dipole is adopted as the excitation source. To reduce leakage, the distance between the positive charge and the negative charge is set to the mesh size of FDTD. At the positions of the two charges, the differential element of electric current densities are
\begin{equation}
	J^\alpha\sqrt{-g}d^3x=(\pm cq,0,0,0)
\end{equation}
where $q$ is the charge. At the midpoint of the two charges, the differential element of electric current density is
\begin{equation}
	J^\alpha\sqrt{-g}d^3x=(0,\frac{dq}{dt}\boldsymbol{l})
\end{equation}
where $\boldsymbol{l}$ is the vector directing from $-q$ to $q$.

Incident wave is incorporated into FDTD  simulation through connection boundaries. The incident electromagnetic fields on connection boundaries can be calculated by Green function method which is introduced in section \ref{sec. green function}.

To obtain far field outside the FDTD domain, we need to integrate on output boundaries using Green function method. The effective electric and magnetic current on output boundaries are:
\begin{equation}
\left\{
\begin{aligned}
	&\boldsymbol{J}(n+1/2)=\boldsymbol{n}\times\boldsymbol{H}(n+1/2)\\
	&\tilde{\boldsymbol{J}}(n)=-\boldsymbol{n}\times\boldsymbol{E}(n)
\end{aligned}
\right.
\end{equation}
where $\boldsymbol{n}$ is the outer normal vector on output boundary, and the variables in parentheses are time steps. We have the difference formulas
\begin{equation}
\left\{
\begin{aligned}
	&\boldsymbol{J}_{,0}(n)=\frac{\boldsymbol{J}(n+1/2)-\boldsymbol{J}(n-1/2)}{\Delta x^0}\\
	&\tilde{\boldsymbol{J}}_{,0}(n-1/2)=\frac{\tilde{\boldsymbol{J}}(n)-\tilde{\boldsymbol{J}}(n-1)}{\Delta x^0}
\end{aligned}
\right.
\end{equation}
where $\Delta x^0=c\Delta t$ is the size of time step. From the law of charge conservation, we have
\begin{equation}
\left\{
\begin{aligned}
	&J^0{}_{,0}(n+1/2)=-\nabla\cdot\boldsymbol{J}(n+1/2)\\
	&\tilde{J}^0{}_{,0}(n)=-\nabla\cdot\tilde{\boldsymbol{J}}(n)
\end{aligned}
\right.
\end{equation}
where the divergence of $\boldsymbol{J}$ and $\tilde{\boldsymbol{J}}$ can be written as the difference of $H$ and $E$ respectively. We have the difference formulas
\begin{equation}
\left\{
\begin{aligned}
	&J^0(n+1)=J^0(n)+\Delta x^0J^0{}_{,0}(n+1/2)\\
	&\tilde{J}^0(n+1/2)=\tilde{J}^0(n-1/2)+\Delta x^0\tilde{J}^0{}_{,0}(n)
\end{aligned}
\right.
\end{equation}

\section{Numerical results}\label{sec. Numerical results}
\begin{figure}[!t]
\centering
\includegraphics[scale=.4]{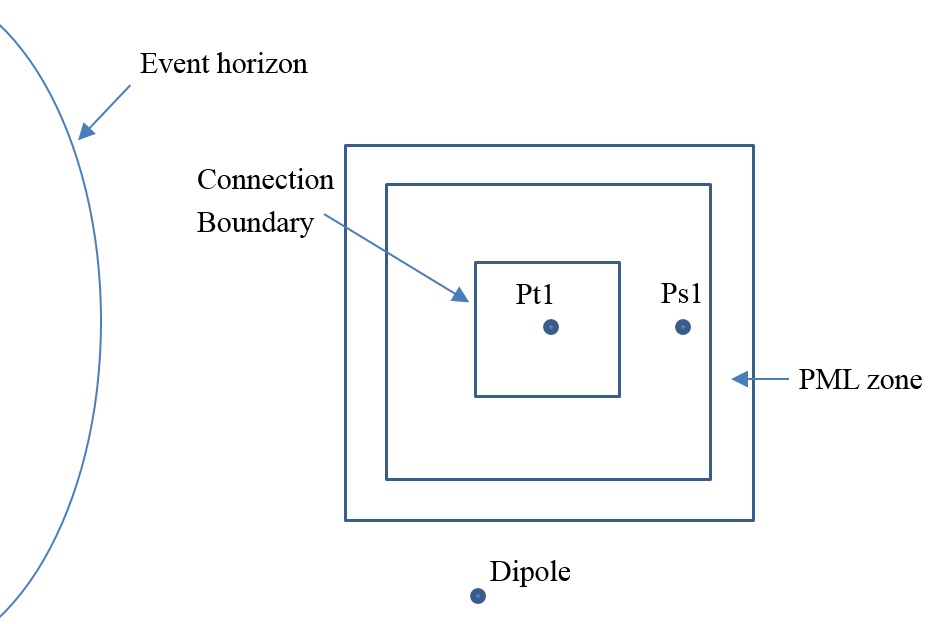}
\caption{Validating the connection boundary} \label{CB}
\end{figure}

\begin{figure}[!t]
      \centering
      \subfloat[Pt1]{\includegraphics[width = .5\textwidth]{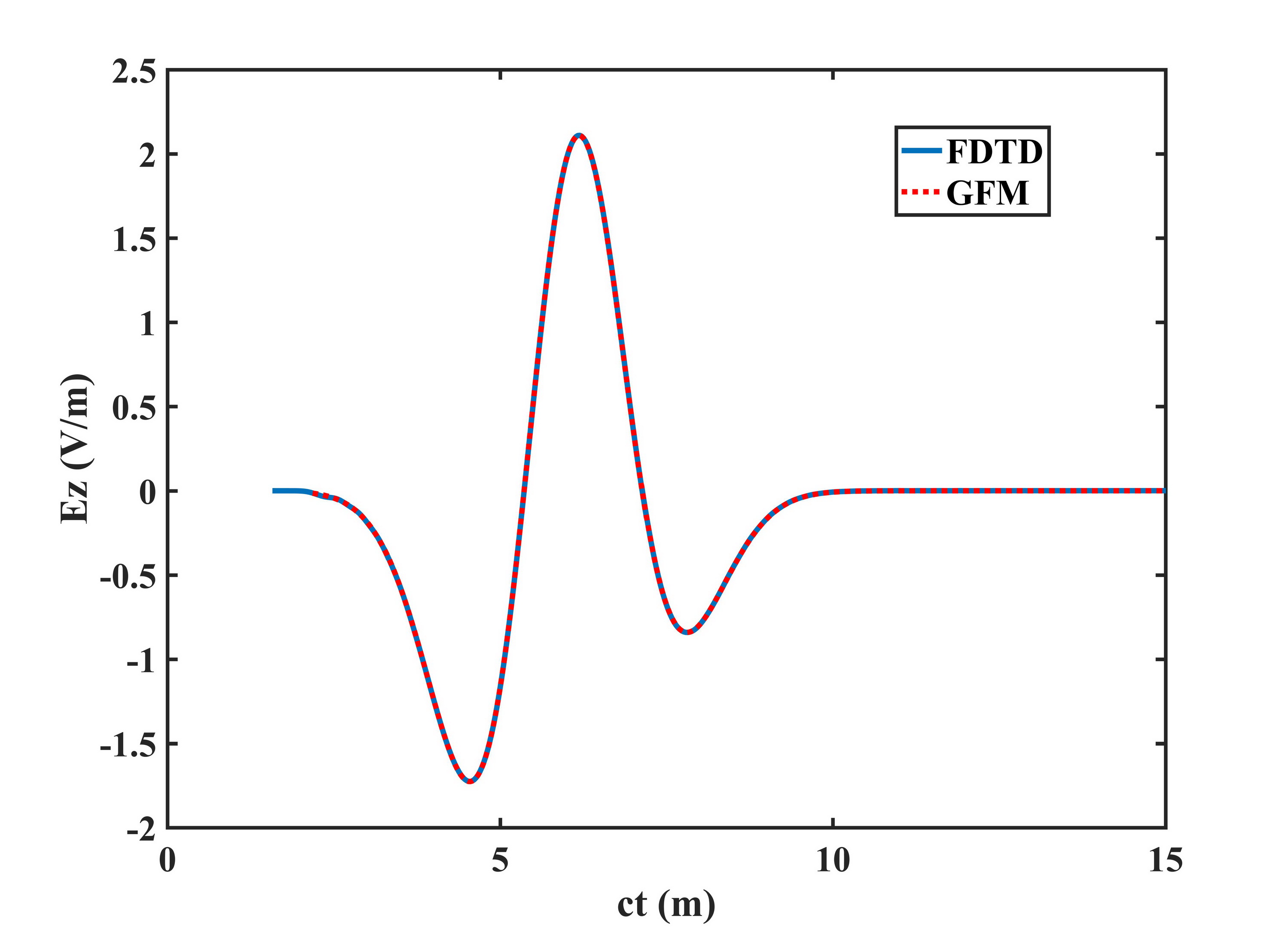}}
      \subfloat[Pt2]{\includegraphics[width = .5\textwidth]{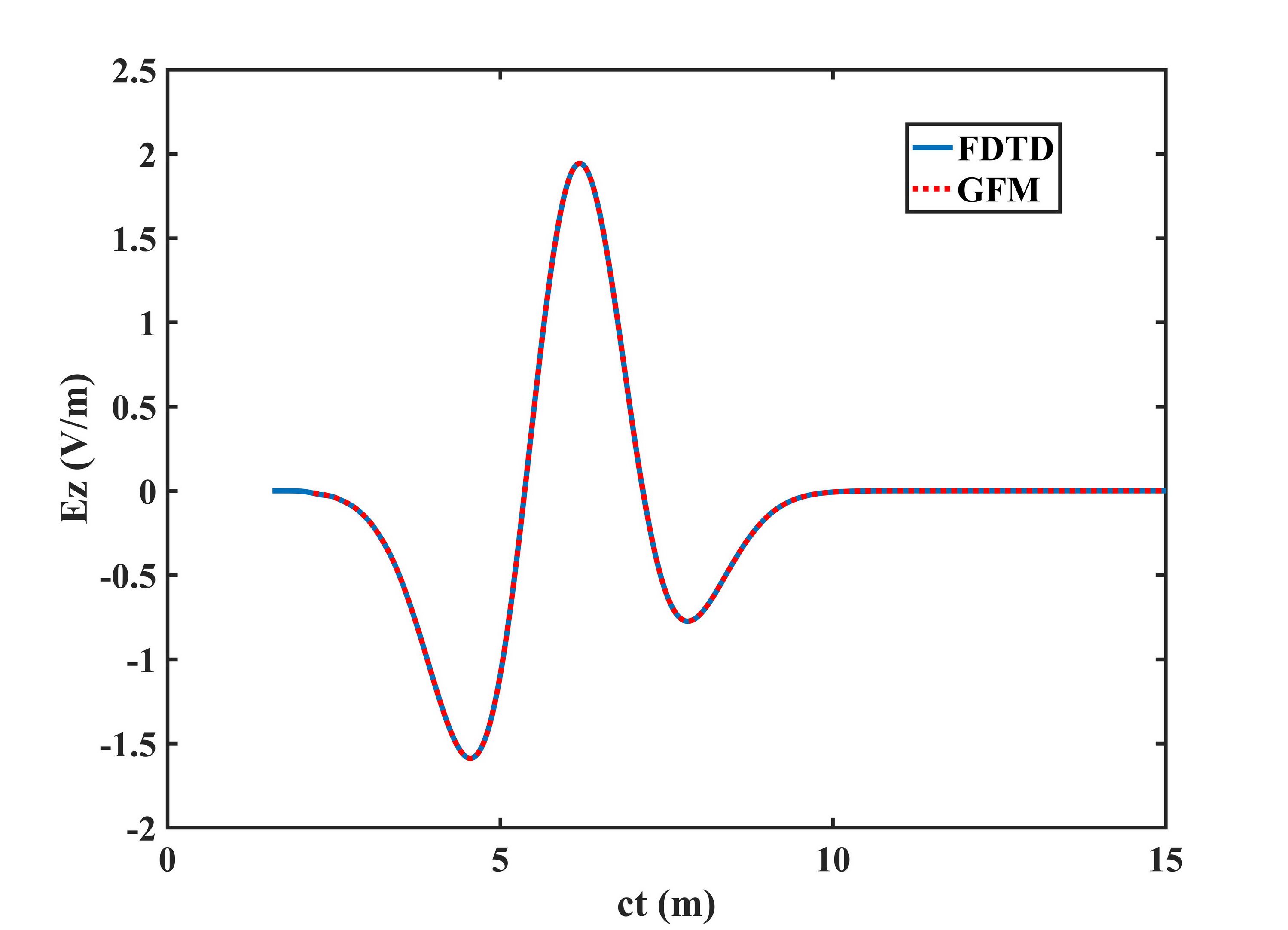}}\\
      \subfloat[Pt3]{\includegraphics[width = .5\textwidth]{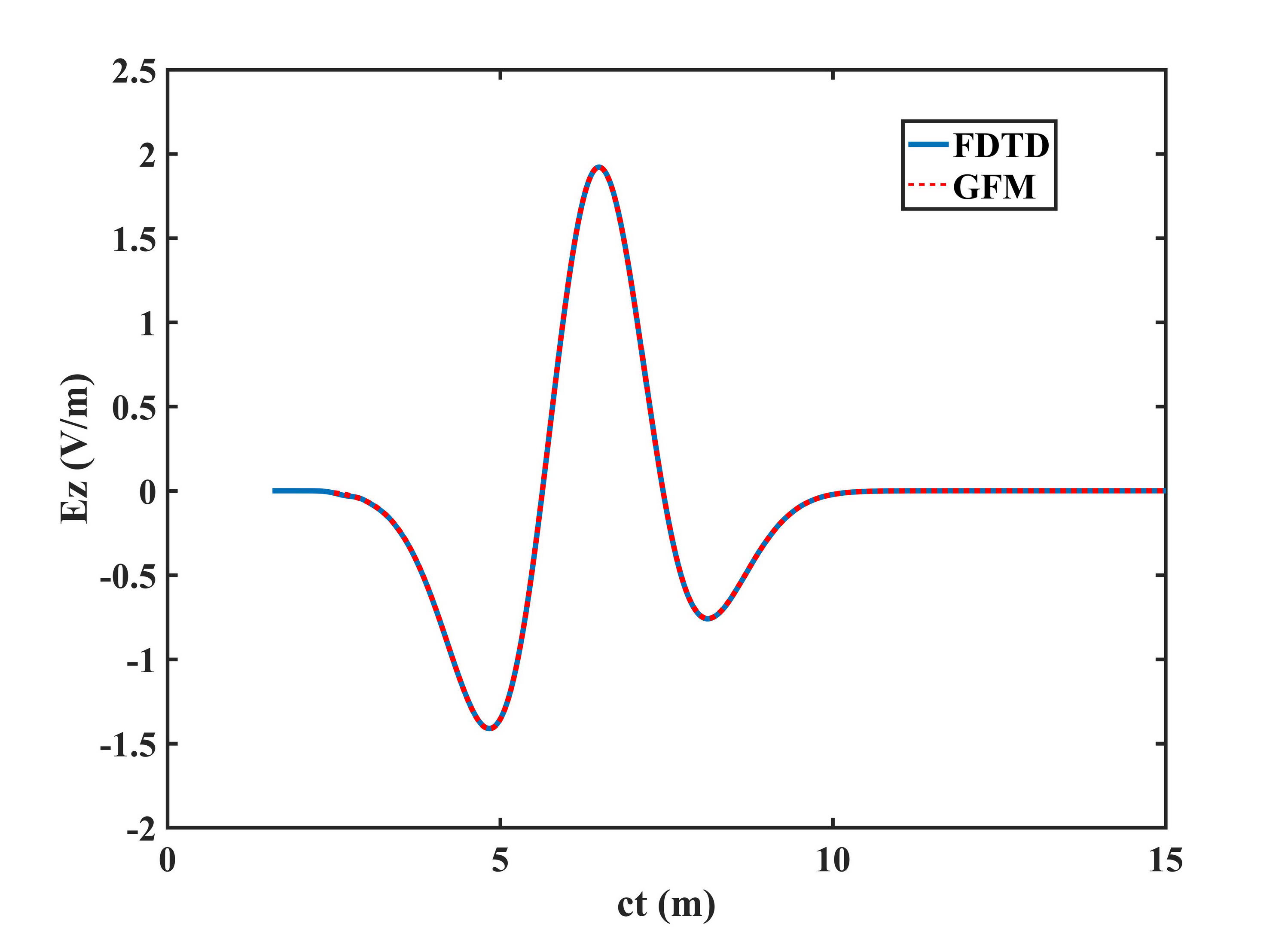}}
      \subfloat[Ps1,Ps2]{\includegraphics[width = .5\textwidth]{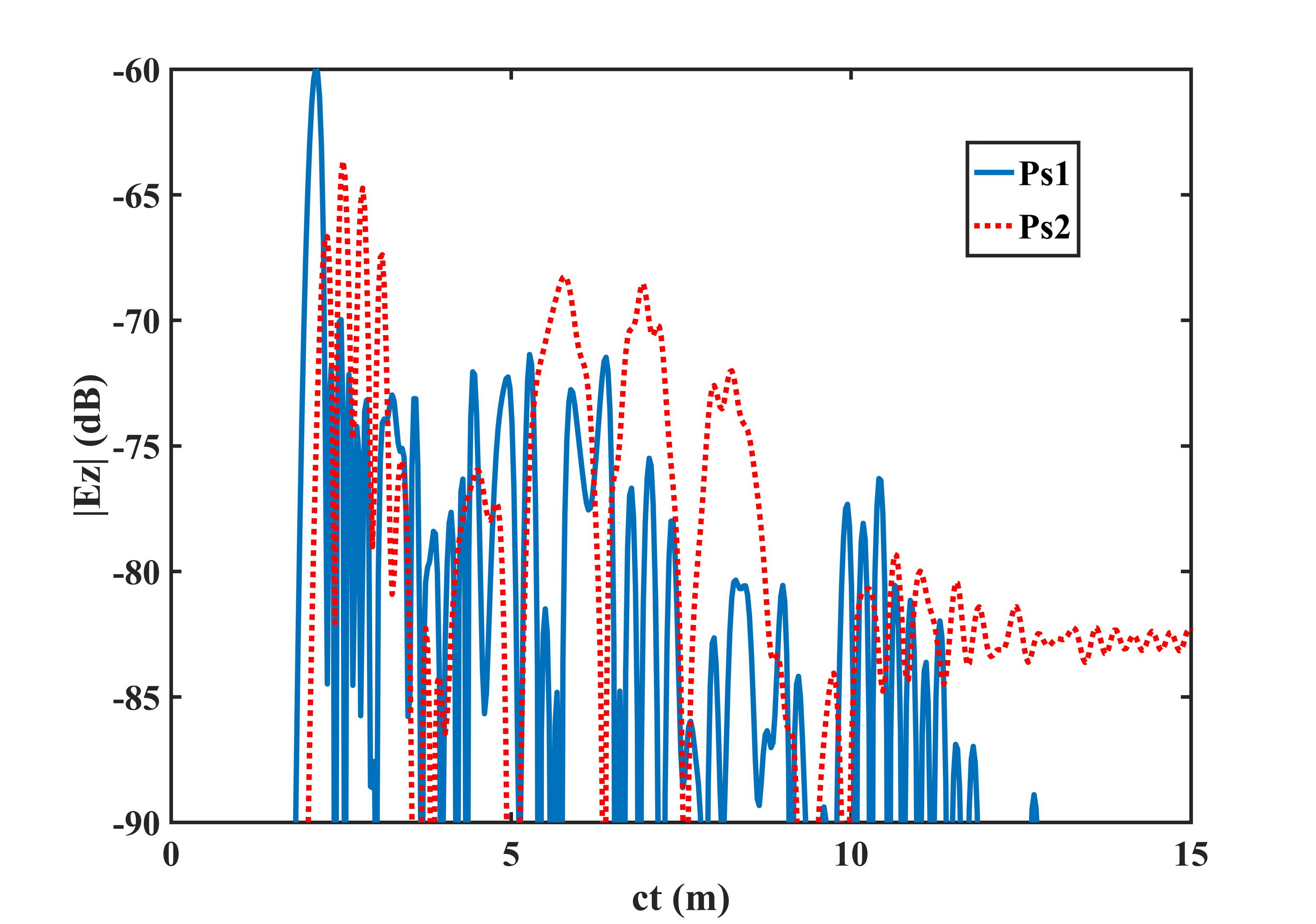}}
      \caption{$E_z$ at Pt1, Pt2, Pt3, Ps1 and Ps2} \label{wave CB}
\end{figure}

The first example is to validate the connection boundary. The Schwarzschild radius is set to 1m. There is no scatterer in FDTD domain (Fig.\ref{CB}). The FDTD mesh size is set to 0.05m. The FDTD domain is: 0.6m to 1.9m in $x$ direction, -0.65m to 0.65m in $y$ direction, and -0.65m to 0.65m in $z$ direction. The number of PML layers is set to 10. The connection boundaries are: $x$=1.25$\pm$0.25m, $y$=$\pm$0.25m, and $z$=$\pm$0.25m. A $z$-directed electric dipole is placed at (1m, -1m, 0m) (the midpoint of the two charges). The waveform of charge is a Gaussian pulse
\begin{equation}\label{Gauss}
	q(t)=A\text{exp}\left[-\frac{1}{2}\left(\frac{t}{\sigma}-4\right)^2\right]
\end{equation}
where $A$=1$\times10^{-6}$C and $\sigma$=3.22ns. The $z$ components of the electric field at five positions are calculated by FDTD method. These positions are Pt1(1.25m, 0m, 0.025m), Pt2(1.4m, 0m, 0.025m), Pt3(1.25m, 0.15m, 0.025m), Ps1(1.7m, 0m, 0.025m) and Ps2(1.25m, 0m, 0.475m). Pt1-Pt3 are located at total field zone and Ps1-Ps2 are located at scatter field zone. The electric fields at Pt1-Pt3 are also calculated by Green function method (GFM). Fig.\ref{wave CB}(a)-(c) show the results at Pt1-Pt3 by the two methods. It demonstrates that the numerical results obtained through both methods match perfectly. This example validates both the FDTD code and Green function code. The results at Ps1 and Ps2 are shown in Fig.\ref{wave CB}(d), in which the values of electric field are in dB: $20\text{lg}(|E_z|/\text{max}|E_{z1}|)$, where $E_{z1}$ is the electric field at Pt1. It demonstrates that the numerical scatter fields are less than -65dB.

\begin{figure}[!t]
\centering
\includegraphics[scale=.5]{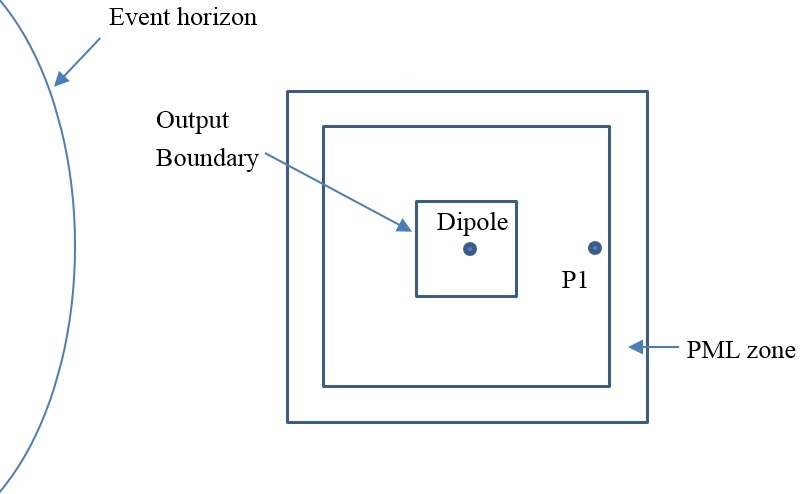}
\caption{Validating the output boundary} \label{OB}
\end{figure}

\begin{figure}[!t]
      \centering
      \subfloat[P1]{\includegraphics[width = .5\textwidth]{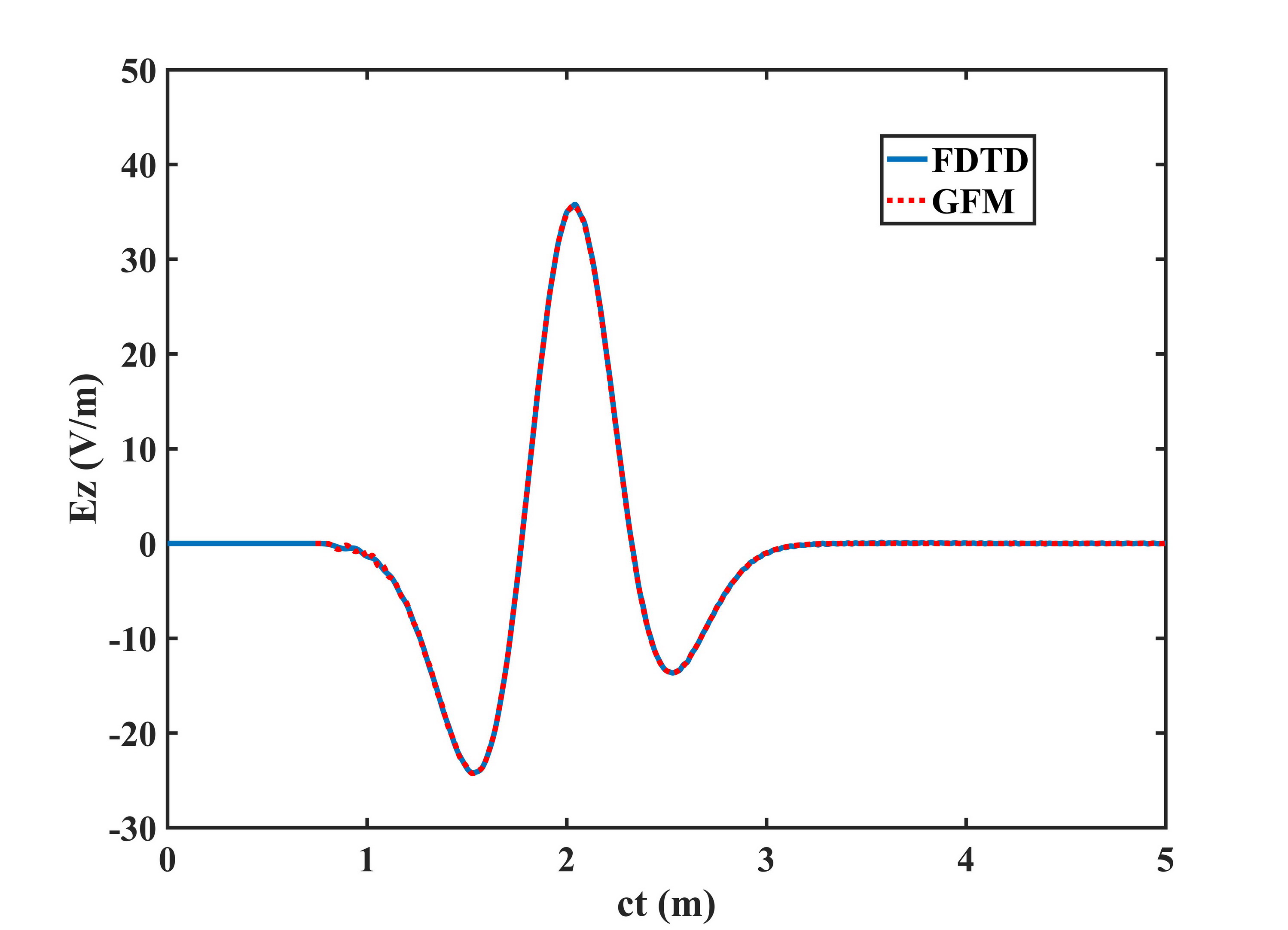}}
      \subfloat[P2]{\includegraphics[width = .5\textwidth]{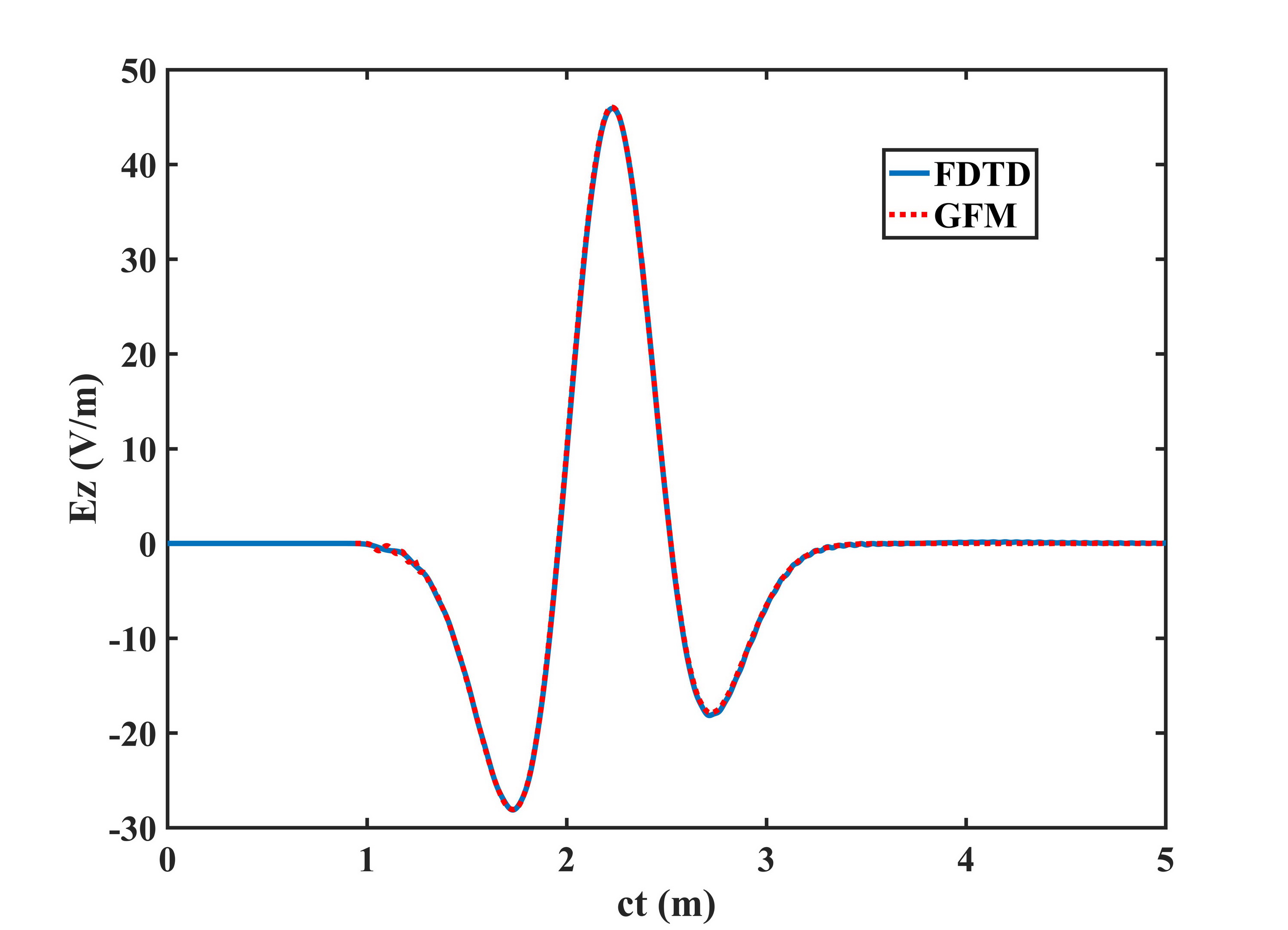}}\\
      \subfloat[P3]{\includegraphics[width = .5\textwidth]{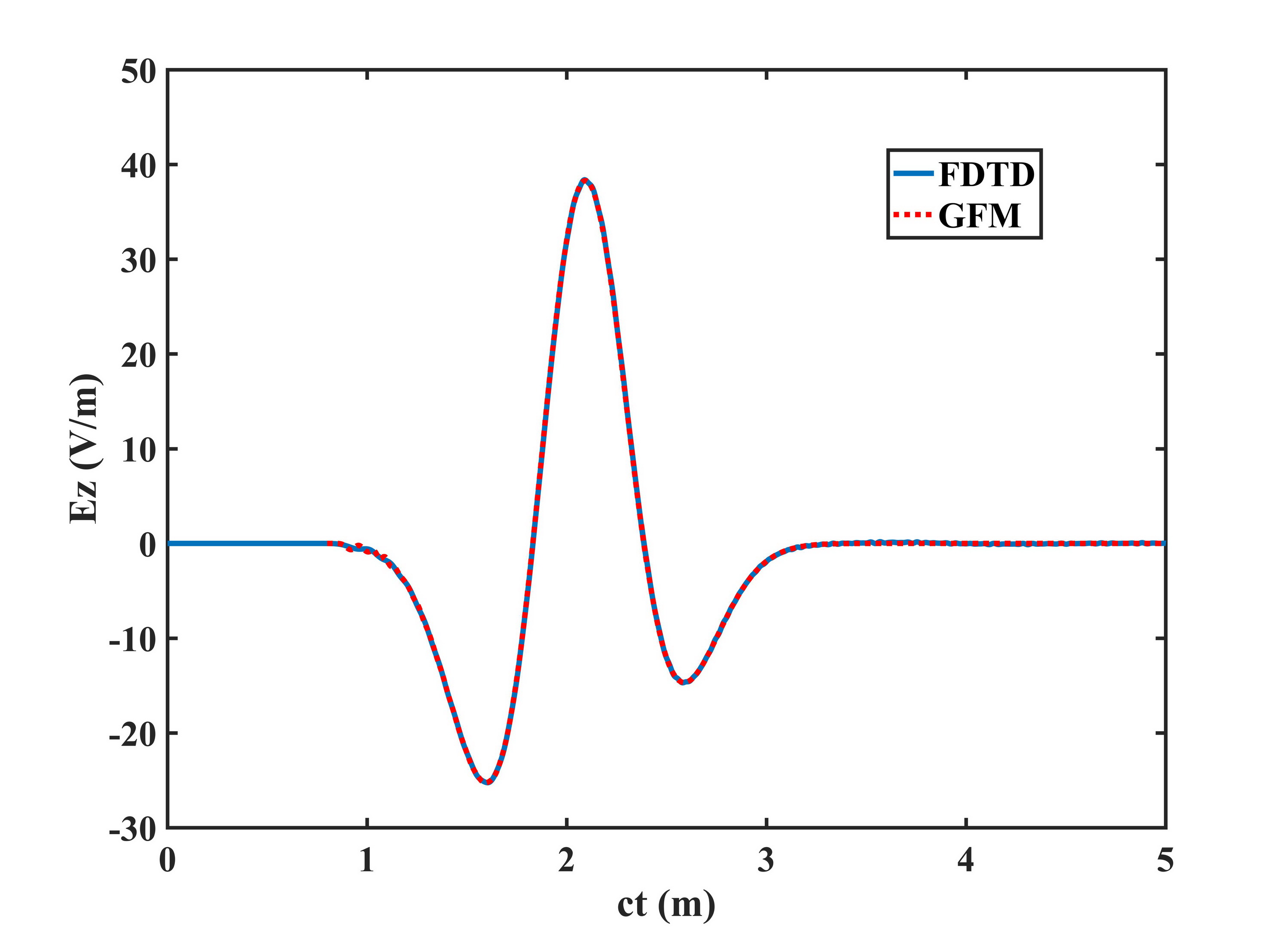}}
      \subfloat[P4]{\includegraphics[width = .5\textwidth]{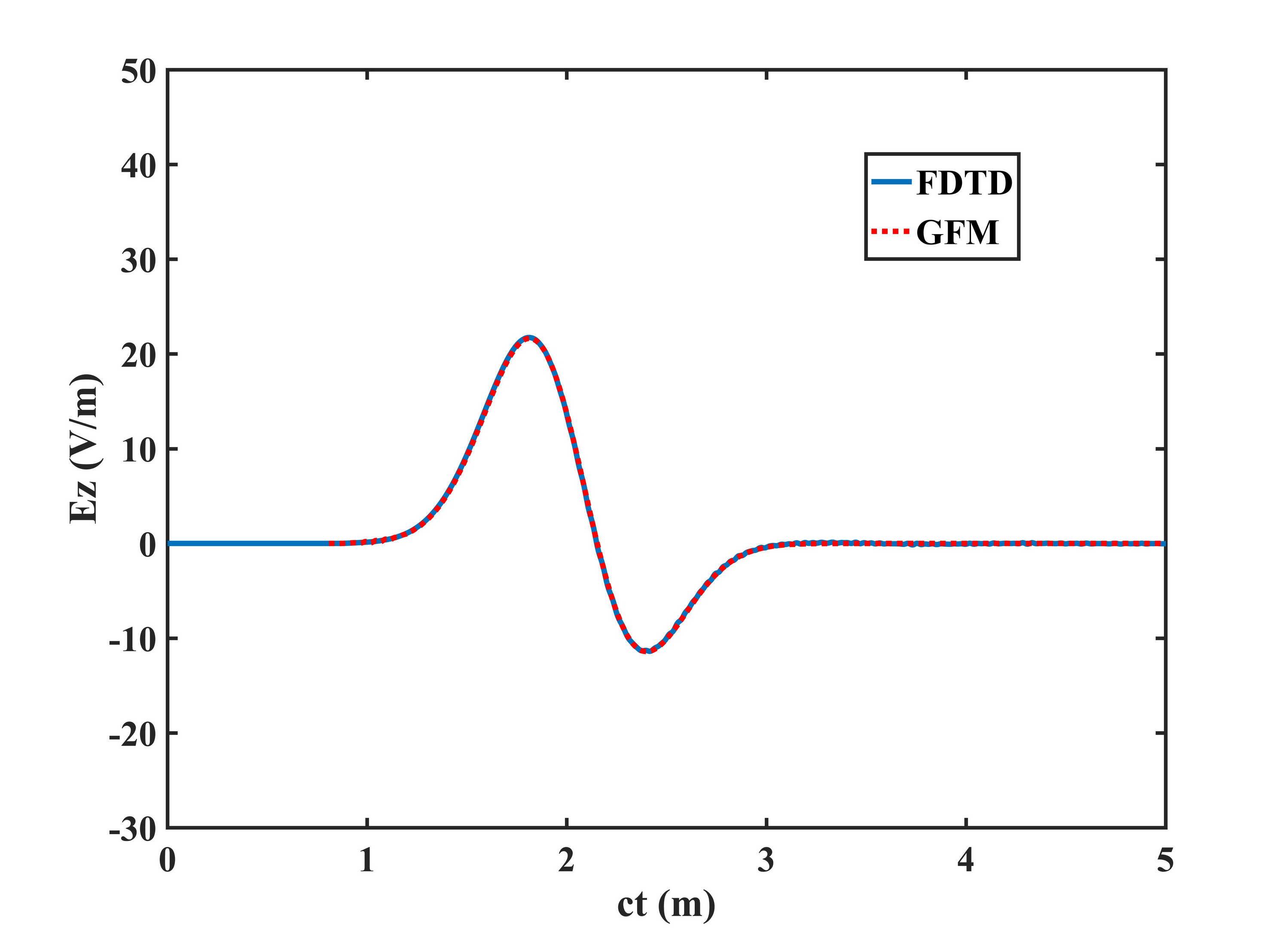}}
      \caption{$E_z$ at P1, P2, P3 and P4} \label{wave OB}
\end{figure}

The second example is to validate the output boundary. The Schwarzschild radius is set to 1m. A $z$-directed electric dipole is placed at (1.5m, 0m, 0m). The waveform of charge is a Gaussian pulse with $A$=1$\times10^{-10}$C and $\sigma$=0.966ns (see Eq.(\ref{Gauss})). The FDTD mesh size is set to 0.01m and the number of PML layers is set to 10. The FDTD domain is: 1m to 2m in $x$ direction, -0.5m to 0.5m in $y$ direction, and -0.5m to 0.5m in $z$ direction. The output boundaries are: $x$=1.5$\pm$0.05m, $y$=$\pm$0.05m, and $z$=$\pm$0.05m (Fig.\ref{OB}). The $z$ components of the electric field at four positions are calculated by FDTD method. These positions are P1(1.98m, 0m, 0.005m), P2(1.02m, 0m, 0.005m), P3(1.5m, 0.48m, 0.005m) and P4(1.5m, 0m, 0.485m). The electric fields at these four positions are also calculated by integrating on output boundaries using Green function method. The results are shown in Fig.\ref{wave OB}. 

\begin{figure}[!t]
\centering
\includegraphics[scale=.5]{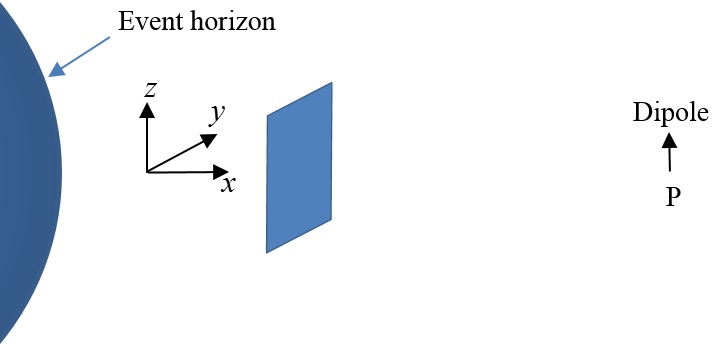}
\caption{Scattering by a thin plate} \label{plate}
\end{figure}

\begin{figure}[!t]
      \centering
      \subfloat[Time domain]{\includegraphics[width = .5\textwidth]{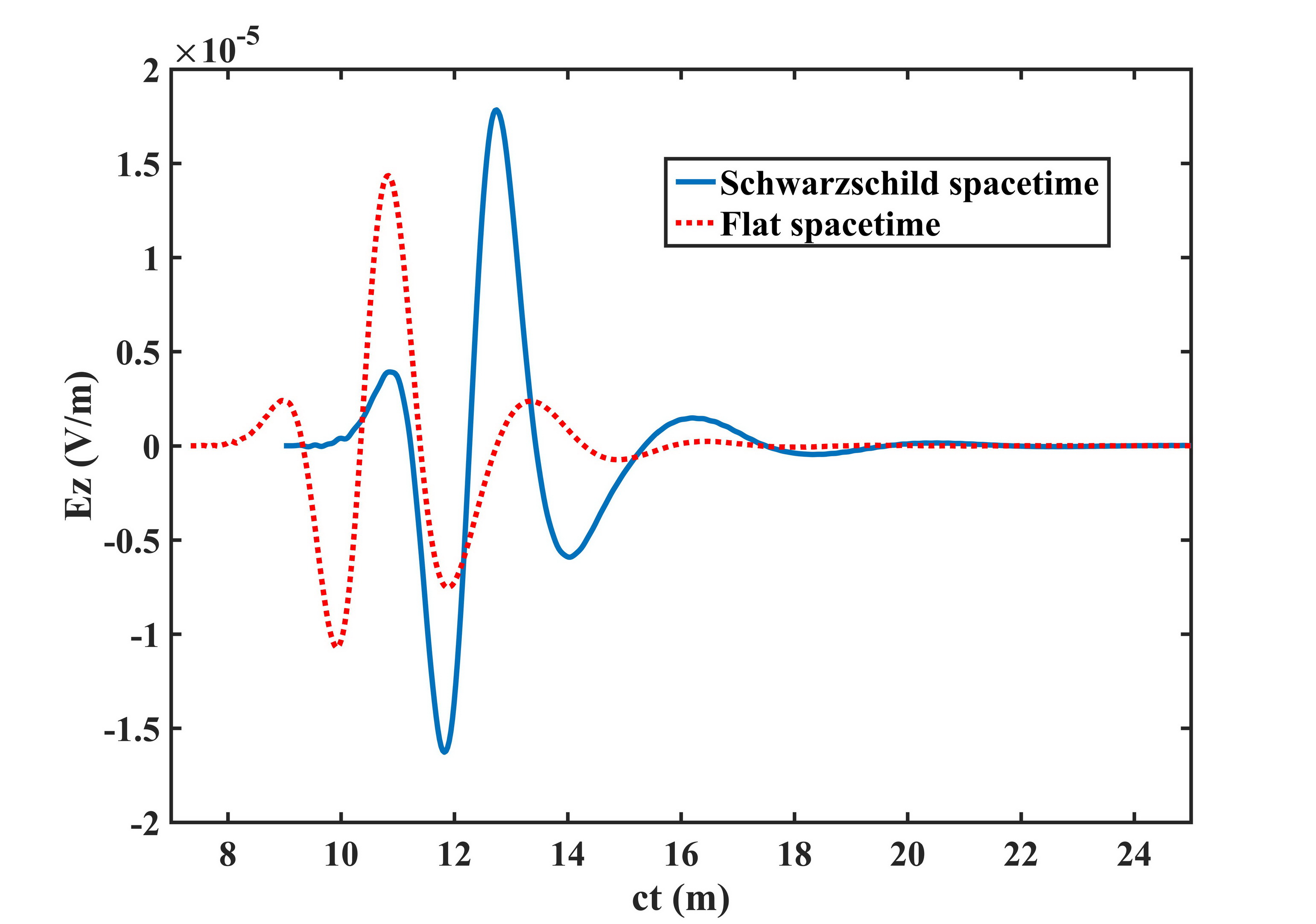}}
      \subfloat[Frequency domain]{\includegraphics[width = .5\textwidth]{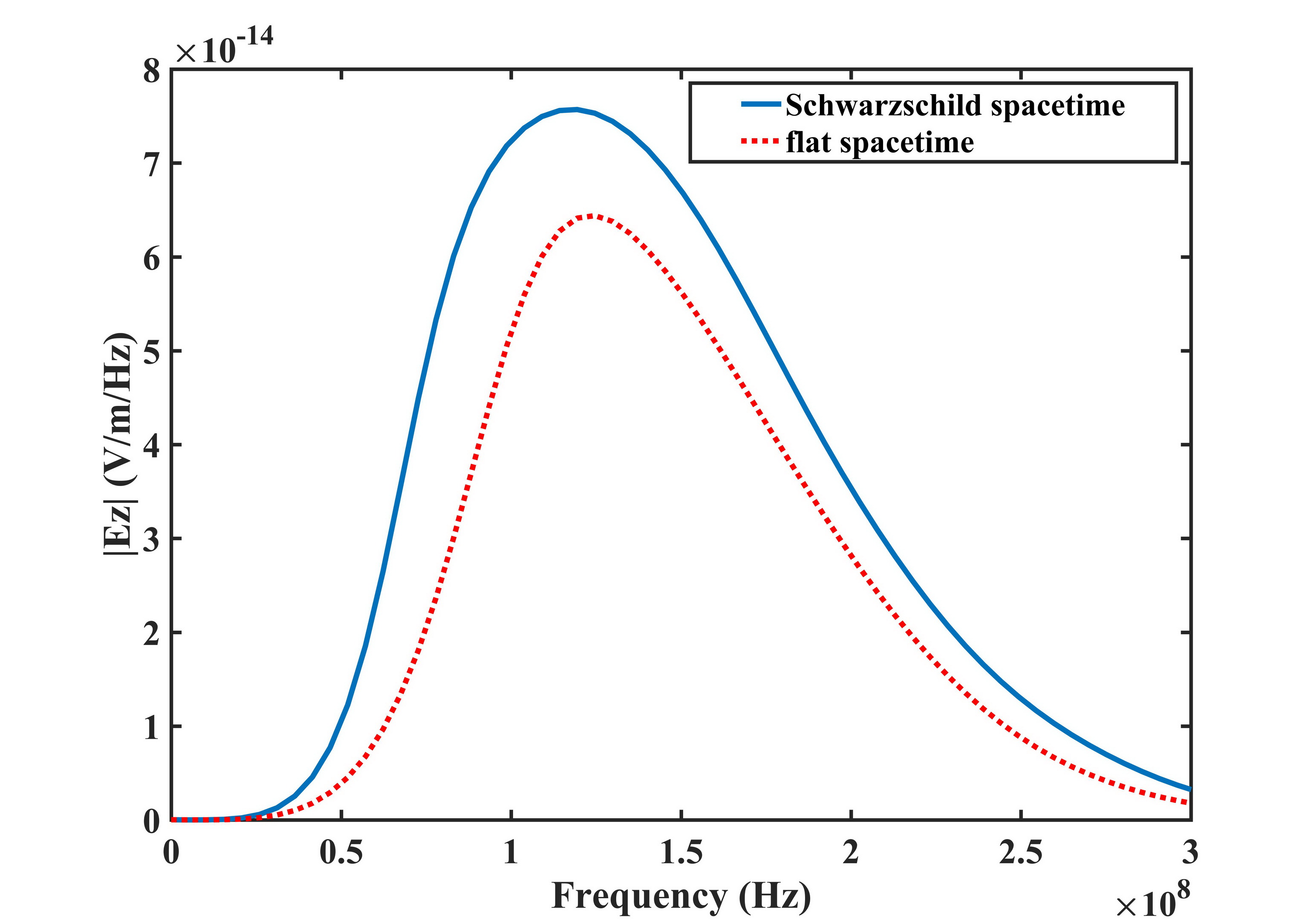}}
      \caption{$E_z$ at P} \label{plate wave}
\end{figure}

The third example is scattering by a thin plate. The size of the thin PEC (Perfectly electric conductor) plate is 1m$\times$1m. It spread out in the plane $x$=3m (Fig.\ref{plate}), and the center locates at (3m, 0m, 0m). The Schwarzschild radius is set to 1m. A $z$-directed electric dipole is placed at P(7m, 0m, 0m). The waveform of charge is a Gaussian pulse with $A$=1$\times10^{-10}$C and $\sigma$=2.415ns (see Eq.(\ref{Gauss})). The FDTD mesh size is set to 0.05m, and the number of PML layers is set to 10. The FDTD domain is: 2.4m to 3.6m in $x$ direction, -1.1m to 1.1m in $y$ direction, and -1.1m to 1.1m in $z$ direction. The connection boundaries are: $x$=3$\pm$0.2m, $y$=$\pm$0.7m, and $z$=$\pm$0.7m. The output boundaries are: $x$=3$\pm$0.4m, $y$=$\pm$0.9m, and $z$=$\pm$0.9m. The $z$ component of the scattered electric field (both in time domain and frequency domain) at P is shown in Fig.\ref{plate wave}. The scattered electric field in flat space-time ($R_S=0$) is also shown in Fig.\ref{plate wave}. From Eq.(\ref{er ur}) we see that the effective light speed is smaller than that in flat space-time. This leads to time delay which is shown in Fig.\ref{plate wave}(a). The inhomogeneity leads to pulse broading in time domain and red shift in frequency domain (Fig.\ref{plate wave}(b)).

\section{Summary and discussion}
FDTD method in in curved space-time is realized by filling flat space-time with equivalent medium. Green function in curved space-time is calculated by solving differential equations. These two methods are incorporated to simulating electromagnetic scattering in Schwarzschild space-time. We validate the FDTD code and Green function code by two numerical examples. The scattering field by a thin plate is simulated by the developed methods.

Eq.(\ref{er ur}) indicates that the effective light speed is smaller as it is closer to the horizon $R=R_S/4$. In previous simulations, the FDTD mesh is uniform, and the mesh size is confined by the lowest effective light speed. In order to save memory, a non-uniform FDTD mesh is feasible. In the simulation, the Green functions between source points and every surface elements on the connection boundaries should be computed, the Green functions between filed points and every surface elements on the output boundaries should also be computed. The computation is very time consuming. To develop the fast algorithm will be our future work.

\section*{Acknowledgment}
This work was supported by the National Natural Science Foundation of China [grant number 61601105].

\end{document}